\documentclass[twocolumn]{aastex631}

\usepackage{type1cm}        
\usepackage{makeidx}         
\usepackage{graphicx}        
\usepackage[bottom]{footmisc}
\let\splitbox\undefined
\usepackage[export]{adjustbox}
\usepackage{xspace}
\usepackage{newtxtext}       %
\usepackage{newtxmath}       
\usepackage{comment}
\usepackage{soul}
\usepackage{amsmath}
\usepackage{subcaption}

\newcommand{\be}{\begin{equation}}
\newcommand{\ee}{\end{equation}}

\newcommand{\bea}{\begin{eqnarray}}
\newcommand{\eea}{\end{eqnarray}}

\shorttitle{Gravitationally-lensed FRBs in NS magnetospheres}
\shortauthors{Dall'Osso et al.}
\begin{document}

\title{GRAVITATIONAL SELF-LENSING OF FAST RADIO BURSTS IN NEUTRON STAR MAGNETOSPHERES:\\ I. THE MODEL}
\correspondingauthor{sim.dall@gmail.com}
\author[0000-0003-4366-8265]{Simone {Dall'Osso}}
\affil{Istituto Nazionale di Fisica Nucleare - Roma 1, Piazzale Aldo Moro 2, 00185, Roma, Italy}
\affil{GSC - Goethe University of Frankfurt, Max von Laue Strasse 1, Frankfurt, Germany}
\affil{Dipartimento di Fisica, Universit\`a La Sapienza, Piazzale Aldo Moro 5, 00185 Roma, Italy}
\author[0000-0003-2810-2394]{Riccardo {La Placa}}
\affil{INAF -- Osservatorio Astronomico di Roma, via Frascati 33, I-00078, Monte Porzio Catone, Italy}
\affil{Dipartimento di Fisica, Universit\`a La Sapienza, Piazzale Aldo Moro 5, 00185 Roma, Italy}
\affil{Dipartimento di Fisica, Universit\`a di Roma ``Tor Vergata'', Via della Ricerca Scientifica, 00133 Roma, Italy}
\author[0000-0002-0018-1687]{Luigi Stella}
\affil{INAF -- Osservatorio Astronomico di Roma, via Frascati 33, I-00078, Monte Porzio Catone, Italy}
\author[0000-0003-0951-8597]{Pavel Bakala}
\affil{Research Centre for Computational Physics and Data Processing, Silesian University in Opava, Bezru\v{c}ovo n\'am.~13, CZ-746\,01, Opava, Czech Republic}
\affil{M. R. \v{S}tef\'anik Observatory and Planetarium, Sl\'adkovi\v{c}ova 41, 920 01 Hlohovec, Slovak Republic}
\author[0000-0001-5902-3731]{Andrea Possenti}
\affil{INAF -- Osservatorio Astronomico di Cagliari, Via della Scienza 5, 09047 Selargius, Italy}
\def\src{1E\,1547.0--5408}
\def\xte{XTE\,J1810--197}
\def\xmm {\emph{XMM-Newton}}
\def\cxo {\emph{Chandra}}
\def\swift {\emph{Swift}}
\def\sax {\emph{BeppoSAX}}
\def\rxte {\emph{RXTE}}
\def\rst {\emph{ROSAT}}
\def\asca {\emph{ASCA}}
\def\flux {\mbox{erg cm$^{-2}$ s$^{-1}$}}
\def\lum {\mbox{erg s$^{-1}$}}
\def\nh {$N_{\rm H}$}
\def\fdo {FRB 20121102A}

\begin{abstract}
Fast Radio Bursts (FRBs) are cosmological sub-second bursts of coherent radio emission, 
whose source~is~still unknown.~To date, the galactic magnetar SGR 1935+2154 is the only astrophysical object known~to emit radio bursts akin 
to FRBs, albeit 
less powerful, supporting suggestions that FRBs originate from magnetars.~Many~remarkable properties of FRBs,~e.g.~the dichotomy between repeaters and one-off sources, and their power-law energy distributions (with typical index $\sim 2-3$), are not well understood yet.~Moreover,~the huge radio power released by the most active repeaters is challenging even for the magnetic energy reservoir~of~magnetars.~Here~we assume that FRBs originate 
from 
 co-rotating hot-spots anchored in neutron star magnetospheres
and get occasionally amplified by large factors via gravitational self-lensing in the strong NS field.~We evaluate the probability of amplification~and~show that (i)~a~power-law energy distribution of events $\propto E^{-(2- 3)}$ is generally expected,~(ii)~all 
 FRB sources may be regarded as repeating, their appearance as one-off 
sources or repeaters being determined by the critical dependence of the amplification probability on the emission geometry~and~source orientation relative to Earth and (iii) the most active repeaters, in particular, correspond to extremely rare and finely-tuned orientations ($\sim$ one in $10^6$), leading to large probabilities of amplification 
which make their bursts frequently detectable.~At the same time, their power release appears enhanced,~typically~by factors $\gtrsim 10$, easing their energy budget problem.
\end{abstract}

\section{Introduction}
\label{sec:1}
Fast Radio Bursts (FRBs) are bright, sub-second transients in the radio sky, with extreme brightness temperatures indicative of a coherent emission mechanism.~Since the earliest discoveries, their dispersion measures (DMs) and dispersive patterns suggested an extragalactic origin \citep[see, e.g.,][]{cordes19,petroff22}, which was later confirmed by the identification of host galaxies and redshift measurements in a few FRBs.~Still eluding us after 15 years of debate, the source of these events must explain many peculiar observational properties, first among which are the huge luminosities and energies of the brightest FRBs:~many events with a precisely determined distance reveal isotropic burst energies $E_{\rm iso} > 10^{39}-10^{41}$~erg, with characteristic millisecond durations. 

A second crucial feature of FRBs is that some of them are observed as repeating sources: since the second detection of \fdo\ in 2015 \citep[][]{spitler14,spitler16}, over fifty repeaters have been identified\footnote{See, e.g., the CHIME repeater catalogue at \url{https://www.chime-frb.ca/repeater_catalog}.}.~This rules out the possibility that all FRBs are produced in catastrophic events, and leads to different scenarios invoking either two distinct populations for one-off and repeating FRBs, or a single population with a very wide range of emission rates and energy budgets.

The number of known FRBs as a function of $E_{\rm iso}$ appears~to 
follow a power-law distribution $\propto E_{\rm iso}^{-\alpha}$, with~a cut-off above $\sim 10^{41}$ erg and $\alpha \sim 2-3$ (e.g.~\citealt{Jam22,Shin23}, \citealt{Lu2020}, \citealt{Gourdji2019}).~Two well-studied repeaters with 
very large
event rate ($\gtrsim 30$ hr$^{-1}$ during active phases) show 
bimodal energy distributions, with the high-energy parts following similar power-laws\footnote{The low-energy part 
resembles a log-normal; the full energy distribution will be addressed in an accompanying paper \citep{LaPlaca2024}.} \citep[][]{li21, zhangetal22}.~Observational 
evidence led several authors to propose that (at least some) FRB sources are related to a cosmic population of magnetars \citep[e.g.][]{Popov2010,Lyutikov2016,Beloborodov2017, Katz17, Margalit2018, petroff22}.~This was~later supported by the detection of FRB-like events from the galactic magnetar SGR 1935+215 \citep{chime20, bochenek20, Dong22, Lu2020, Giri23},
The coherent radio emission might be produced inside the magnetosphere
(e.g.~\citealt{Lyutikov2016,Lyutikov2017}; \citealt{LyuRaf19}; \citealt{Lu2022}, \citealt{Zhu23}).~
In fact, it has been shown \citep{lyu21a,Lyu21b, qu22, Lyu23} that FRB radio~waves can travel through~neutron star (NS) magnetospheres suffering little/no absorption in a~super-strong $B$-field, or if they travel (i) at a small angle ($\lesssim 0.1-0.3$ rad) relative to the local B-field and/or (ii) in a relativistically-moving plasma ($\Gamma \gtrsim 10$).~Both conditions~can be met on~the open field-lines of the magnetosphere, regardless of the B-field strength.~Within the closed field-line region, where 
neither condition is~met, a strong B-field may prevent absorption only up to~a maximal distance $R_{\rm abs} \approx 3.5 \times 10^8$ cm~$\mu^{1/2}_{33}/L^{1/4}_{42}$, 
due to its steep radial decline.~
Here $L$ is the isotropic-equivalent luminosity\footnote{In this work we adopt the shorthand $Q_x=Q/10^x$ for various quantities expressed in cgs units, unless otherwise stated.} of the FRB wave and $\mu$ the NS magnetic moment \citep[][]{beloborodov21}.~Waves 
travelling through the closed-field-lines may thus be absorbed at large distances from the NS, provided the spin period is long enough ($ P > 0.1$ s) for the closed field-line region to extend beyond $R_{\rm abs}$ \citep[cf.][]{Lyu23}.~Such factors may introduce a natural selection on the orientation and properties of observable FRB sources\footnote{Additionally, it can lead to a frequency-dependent temporal broadening of FRB pulses \citep{soba22}.}.

In this work, we argue that the large energies and power-law energy distribution of bright FRBs admit a straightforward interpretation, in terms of gravitational (self-)lensing of flares produced in the magnetospheres of highly magnetic NSs, at a small distance from the surface.~At the same time, this scenario naturally predicts the existence of repeaters and of one-off FRBs, as a result of geometry/orientation effects in a single population of otherwise homogeneous sources.~Our scenario is agnostic about the emission mechanism; it only requires~a magnetospheric origin.

\section{Preliminary remarks}
Most applications of gravitational lensing make use of the weak deflection limit (WDL), owing to the characteristically large distance, $R$, between celestial bodies~\citep[e.g.][]{aaa06, jetzer10}.~The lens amplification $a$ rises~sharply as the source gets closer to the caustic line, by which 
we refer to the continuation of the line-of-sight behind the lens, with respect to the observer. Here, instead, we consider lensing in the vicinity of an NS, as approximated by a Schwarzschild lens.~Since the emitting source lies close to the lensing body and the latter is a compact object, strong field gravity effects will come into play \citep[$R \lesssim 200 GM/c^2 = 200 R_g$;][]{Bakala2023};~we denote the regimes of large and small angular separations ($\theta$) from the caustic line, respectively, as weak lensing regime (WLR) and extreme lensing regime (ELR).


~In the ELR the amplification scales roughly as~$a \propto \theta^{-1}$~and, for point-like sources, it is only limited by diffraction to a maximum amplification\footnote{This expression holds exactly in the WDL and to within one order of magnitude in the ELR \citep[see, e.g.][]{Deguchi1986a,Deguchi1986b,Nambu2013}.}
 $a_{\rm diff} \simeq 4 \pi^2 R_g/\lambda \approx 
4\times10^5 (M/ M_\odot) \nu_{\rm GHz}$ for radiation at wavelength $\lambda$, or frequency $\nu$ (in GHz),  and a lens of mass~$M$. 
With finite-size sources, of linear scale $\ell$ and angular size~$\theta_s = \ell/R$ relative to the center of the lens, the amplification saturates as the angular separation is reduced below~$\theta_s$. In these cases, the source finite size may become the main factor limiting the amplifications to values lower than $a_{\rm diff}$.  

The amplification provided by
 NS self-lensing 
naturally reduces the energy requirements of bright FRBs without invoking highly relativistic motion.~Moreover, as we will show, 
it offers a key to explaining the dichotomy between repeaters and one-off FRBs.~Indeed, under the assumption that the seed events occur, in each source, at a fixed magnetospheric location co-rotating with the NS, 
we find that orientation/viewing angle effects naturally lead to such a dichotomy, even for a source population made entirely of identical repeaters.~This stems from the 
fact that 
in a random sample most orientations 
have extremely small probabilities of achieving large amplifications, while at the same time 
there exists a minute range of finely-tuned orientations 
which guarantee large amplification probabilities.~In this picture, self-lensing can also ease problems with the 
(time-integrated) energy budget of the most frequent repeaters, e.g. FRB 20121122A and FRB 20201124A, which are known to challenge even the energy reservoir of magnetars magnetospheres 
(\citealt{li21}, \citealt{zhangetal22}).

Gravitational self-lensing naturally produces a 
truncated 
power-law probability distribution of the amplifications, $P(a) \propto a^{-(2-3)}$, with the index determined by the 
location of the emitting source in the magnetosphere.~Accordingly, the same power-law will shape the energy distribution of the amplified events, 
(almost) regardless of the seed energy distribution.~The latter may become apparent in the low energy tail of the observed distribution, if it happens to be above our 
detection threshold.~Interestingly, 
\begin{figure*}
\centering
\begin{subfigure}{0.495\textwidth}
\centering
\includegraphics[height=0.62\linewidth, inner]{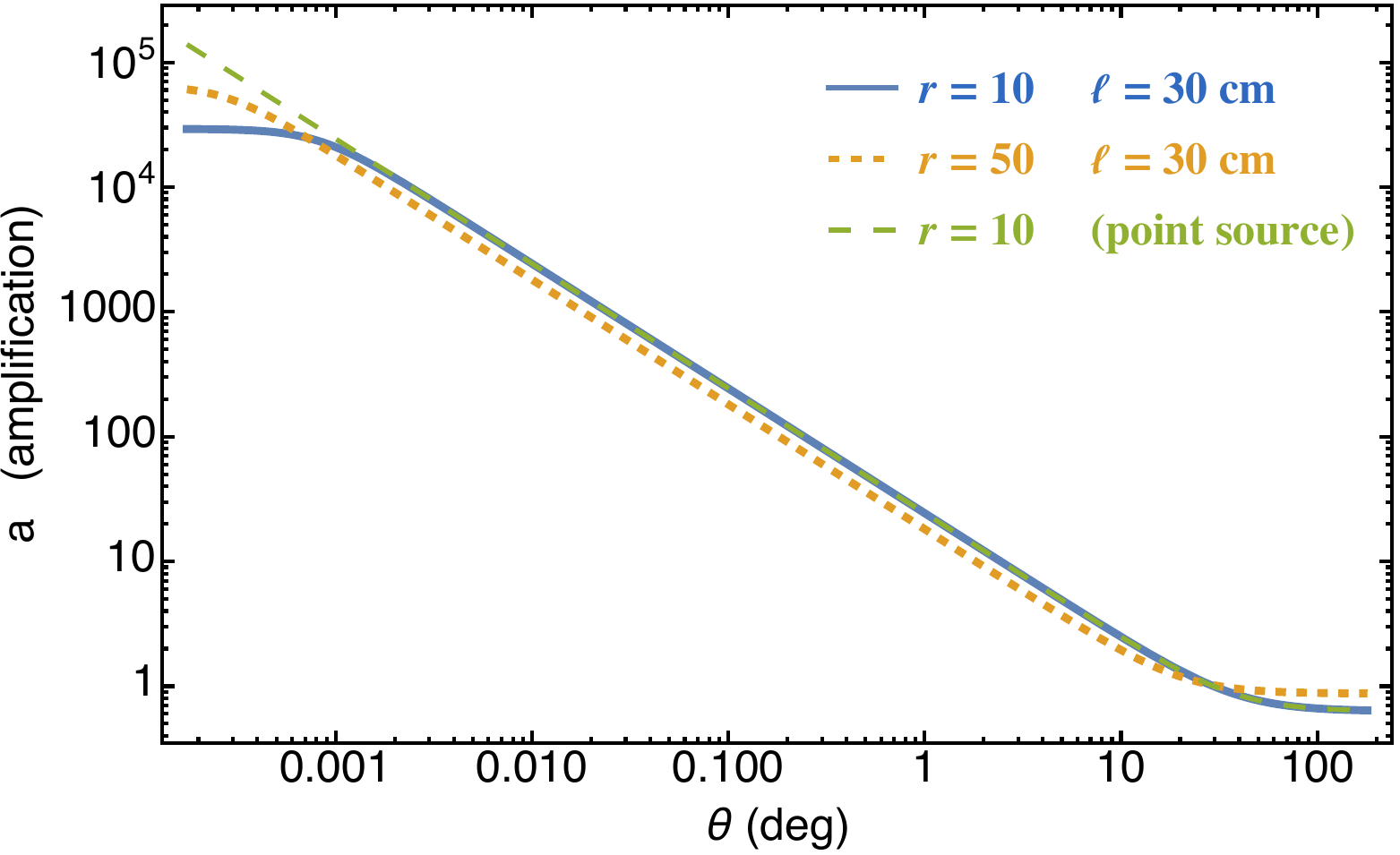}
\begin{minipage}{0.1cm}
\vfill
\end{minipage}
\end{subfigure}
\hfill
\begin{subfigure}{0.495\textwidth}
\centering
\includegraphics[height=0.62\linewidth, right]{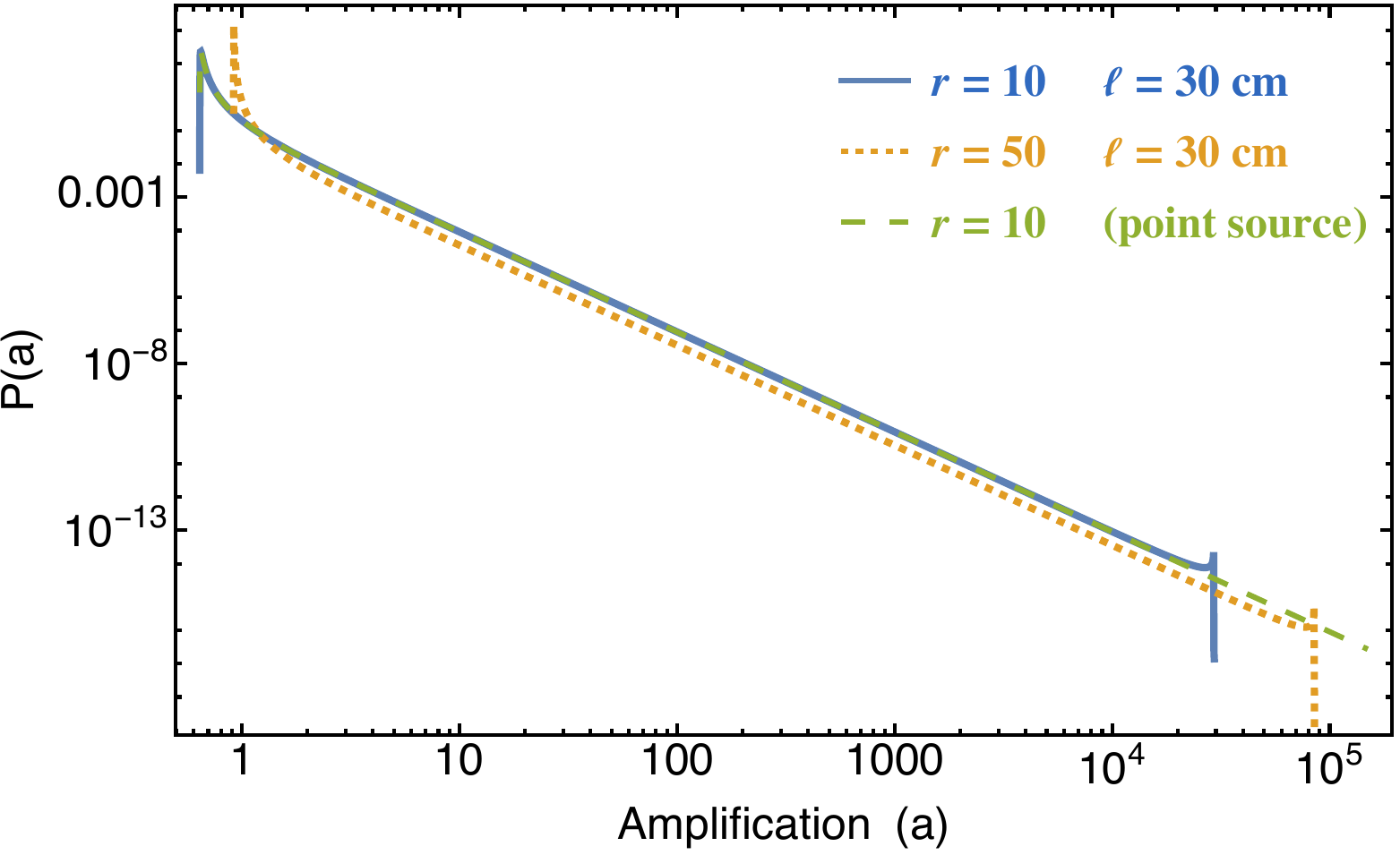}
\begin{minipage}{0.1cm}
\vfill
\end{minipage}
\label{fig:magn-b}
    \end{subfigure}
\caption{{\it Left Panel:} The amplification $a = \mu g^4$ vs. angle $\theta$ from the caustic line, for a 1.4$M_\odot$ NS and (i) a point-like source~at an emission radius $r=10$ (green, dashed), (ii) a finite-size source of linear size $\ell=30$~cm at $r=10$ (blue), (iii) the same finite-size source at $r=50$ (orange). The dashed green and solid blue curves overlap at an angular distance $\gtrsim \theta_s = \ell /r$,~where the source size plays no role.~The orange curve reaches larger amplifications than the blue one because $\theta_s$ is smaller at larger radii, hence~the source gets closer to the caustic line before its size becomes relevant.~{\it Right Panel:} Probability distribution $P(a)$ for each curve of the left panel.~The peaks at low amplification ($a < 1$, due to gravitational redshift) indicate the large probability for random bursts to occur far from the caustic line.~Such~peaks occur, in the blue and orange curves,~at $a \approx 0.65$ and 0.95, respectively.~For reference, the blue and orange curves have probabilities $\lesssim 0.2$\% and $\lesssim 0.07\%$, respectively, of having $ a > 5$.}
\label{fig:regimes}
\end{figure*}
the~rare Giant Pulses (GPs) observed in some radio pulsars also show a similar power-law energy distribution \citep[e.g.][]{mckee19,Bilous2022,Lin2023}, in contrast to typical pulses which generally 
follow broad log-normal distributions \citep[]{Cairns2001,BurkeSpolaor2012,Mickaliger2018}:~a 
possible link between FRBs and GPs was already proposed by \citet{cordes16} and \citet{Connor2016}, motivated by their very large brightness temperatures.

This is the plan of the paper:~sec.~\ref{sec:Pdia} introduces general features of gravitational self-lensing that~will leave an imprint on FRB observed properties, presents the weak and extreme lensing regimes, and calculate the overall effect of lensing~when the whole NS magnetosphere can produce localised~bursts~at random times.~Sec.~\ref{sec:rings} extends this calculation to a different configuration, assuming a small magnetospheric active~region, stably anchored to the NS, and compares the results.~Sec.~\ref{sec:obsvsint} describes the differences between the observed (apparent) and actual source properties that are caused by gravitational self-lensing.~In sec.~\ref{sec:implications} we outline the main astrophysical implications of our model, and sec.~\ref{ssec:popstudy} shows that a subset of the cosmic population of NS can account well~for the observed number of FRBs, both repeating and~non-repeating.

\section{Probability distribution of amplifications in active magnetospheres}
\label{sec:Pdia}
In our proposed scenario, FRB emission is generated by~a small hot-spot in the NS magnetosphere, at a small~distance ($ R < 100 R_g$) from the star.~The emission may be beamed by a factor $f_b \ll1$, in a direction assumed to be random~at each event (more details in Appendix \ref{app:pdevpda}).~ 
If~the emission~occurs close to the caustic line, and beamed in a favorable direction, it gets amplified by a large factor $a$,~which~greatly enhances its detectability.~If little or no amplification~takes place, though, FRB detection is hampered,~and only occurs if the intrinsic luminosity of the event~is~large enough.~The bolometric amplification $a$, results from the geometrical magnification given by gravitational lensing, $\mu = \left| (\beta / \theta) d\beta/d\theta \right|$, modified by the energy decrease due to the NS gravitational redshift, $g^4 = (1-2R_g/R)^2$.~Here $\theta$~and   
$\beta$ indicate, respectively, the angular distance of the~source, and of its image~in the sky, from the caustic.~For a Schwarzschild lens, the~amplification of a point-like source in the ELR scales with~$\theta$,~and with the radial distance from the star, $r$, as \citep{Bakala2023}
\be
\label{eq:apointlike}
a(r,\theta) = 2  \displaystyle\frac{\left[1 -  \displaystyle \left(2/r\right)^{0.85}\right]^{1.35}} {r^{1/2} \theta} \equiv 2 ~\displaystyle \frac{f(r)}{r \theta} \, ,
\ee
where $r = Rc^2/GM = R/R_g$ is dimensionless,~$R$ being~the distance in physical units and $R_g$ the gravitational radius,~and $f(r) = r^{1/2} \left[1 -  \displaystyle \left(2/r\right)^{0.85}\right]^{1.35}$. 

At large angular separations the WLR~is reached, in which $a$ has a much flatter scaling with~$\theta$.~In~order to derive a simple, analytical approximation that could incorporate all lensing regimes, we generalized eq.~(\ref{eq:apointlike}) as 
\be
\label{eq:aweaklens}
a(r,\theta) = 2 ~\displaystyle \frac{f(r)}{r \theta} ~ \left[1 + \displaystyle \left(\frac{\theta}{\theta_{\rm we}}\right)^{S}\right]^{1/S}\, ,
\ee
Here $\theta_{\rm we}$ and $S$ depend on $r$ and describe, respectively,~the angle at which the transition from extreme to weak lensing occurs, and the sharpness of that~transition.~They were~derived by approximating the exact expression for $a$ as described in 
appendix~\ref{app:A}.%

With these provisions eq.~(\ref{eq:aweaklens}) gives an accurate fit to the full solution of the Schwarzschild lens equation in {\it any} regime, while the much simpler eq.~(\ref{eq:apointlike}) gives the correct limit in the ELR (Fig.~\ref{fig:regimes}).~The above definitions allow us to calculate the probability distribution for $a$, under differerent assumptions for the geometry of the emission region.

\subsection{Point-like sources} 
\label{sec:magsphere}
The simplest possible situation is that of point-like sources going off at any time and at random positions in the NS magnetosphere, up to some maximum radial distance $R_{\rm max}$.~The probability that their emission gets amplified by a factor~$a$ will depend~on the ratio between the volume within which that particular~$a$-value can be achieved, and the whole available volume ({\it i.e.}~the region with $R \le R_{\rm max}$).~For given $R$ and $\theta$ the volume element within $dR$ and $d\theta$ is $dV(\theta, R) = 2 \pi R^2 \sin \theta d\theta dR$, 
from which the %
probability distribution for $a$ in the spherical shell between $R$ and $R+dR$ is obtained
\begin{eqnarray}
\label{eq:Pda-spherical}
P(a; r)da & = & \displaystyle \frac{dV[\theta(a)]}{4 \pi R^2 dR} = \frac{\sin \theta \left(d \theta/da\right)}{2}~da \nonumber \\
 & \simeq & \displaystyle \frac{f(r) \sin \left[2 f(r)/a\right]}{ a^2}~da  \, .
\end{eqnarray}
The last step in eq.~(\ref{eq:Pda-spherical}) has the well-know form $P(a) \propto a^{-3}$: it is valid in the ELR, where $d \theta / da$ is given by~eq.~(\ref{eq:apointlike}). In practice this means that, e.g., 
at $r=10$ %
the~$a^{-3}$~scaling holds for $\theta \le \theta_{\rm we}\approx \pi/6$, corresponding~to~$a \gtrsim 0.93$.~At~larger~$\theta$, hence lower amplification, 
the marked flattening of the scaling ($a \sim$~const.) 
in the WLR leads instead to a pronounced~steepening of $P(a)$ at low amplifications (Fig.~\ref{fig:regimes}, right panel).

Finally, by integrating eq. (\ref{eq:Pda-spherical}) 
over the volume of the active magnetosphere, the global $P(a)$ is obtained.~To this 
aim, we assumed that emission may occur at different radii following a probability distribution $P(r)\propto r^{-k}$, and set:~(i) the inner edge at $r_{\rm in}\approx 10$, to avoid the inner region likely occulted by the NS (see Appendix \ref{app:A}),
(ii) the outer edge at $r_{\rm max} = 50$.~Note that the result depends only weakly on the value of the outer edge, owing to the fast decrease of the lensing probability with growing distance from the lens.
Regardless of the value of $k$, we find that $P(a)$ maintains~the $a^{-3}$ scaling at large amplifications (Fig.~\ref{fig:radial}, left panel), while the radial extension of the active magnetosphere only affects 
the low-amplification end (see also next~sub-section).~This too is a result of the fast decrease of the lensing probability with growing~$r$, which gives a growing fraction of weakly or non-amplified~events. 
 
\subsection{Finite-size sources} 
\label{sec:finitesize}
The finite dimension ($\ell$) of actual sources will of course limit the maximum achievable amplification.~In particular, $a$ is expected to saturate when the angular 
separation becomes comparable to, or smaller than, the source angular size relative to the lens, {\it i.e.} $\theta \lesssim \theta_s \sim ~\ell/R$.~We  parameterize this effect~as
\be \label{eq:afinitesize}
a_{\theta_s} (r, \theta) = \frac{2f(r)}{ r \theta_s \left[1+\left(\displaystyle \frac{\theta}{\theta_s}\right)^p\right]^{1/p}} \, .
\ee
The saturation occurs in the ELR, as long as the emitting~region is relatively small ($\ell < 0.1 R$).~Hence, we
set the smoothness parameter $p = 3$ in order to match eq.~(23) in \citet{Bakala2023}, which is exact for finite-size sources in the~ELR. 
We incorporated this saturation by adding the square parenthesis of eq.~(\ref{eq:afinitesize}) in the denominator of eq.~(\ref{eq:aweaklens}) as detailed in Appendix \ref{app:A}. The general relation $a(\theta)$ for finite-size sources thus derived can~be roughly described as a double broken power-law (see, e.g., Fig.~\ref{fig:regimes}, left panel).~At~small angular separations, $\theta < \theta_s$, a flat segment with $a \simeq a_{\rm max}=2 f(r)/(r \theta_s)$ is followed, in~the range $\theta_s < \theta < \theta_{\rm we}$, by a power-law~decay $a \propto \theta^{-1/3}$.~After the transition to the weak WLR, at~$\theta \ge \theta_{\rm we}$, it turns to another shallow segment~where $a \rightarrow {\rm const} $ as $\theta \rightarrow \pi$.
The right panel of of Fig.~\ref{fig:regimes} illustrates the $P(a)$ distributions corresponding to each of the cases shown in the left panel (see~caption for details).
\begin{figure*}[ht]
\centering
\begin{subfigure}{0.495\textwidth}
\centering
\begin{minipage}{0.9cm}
\vfill
\end{minipage}
\includegraphics[width=1.01\linewidth, inner]{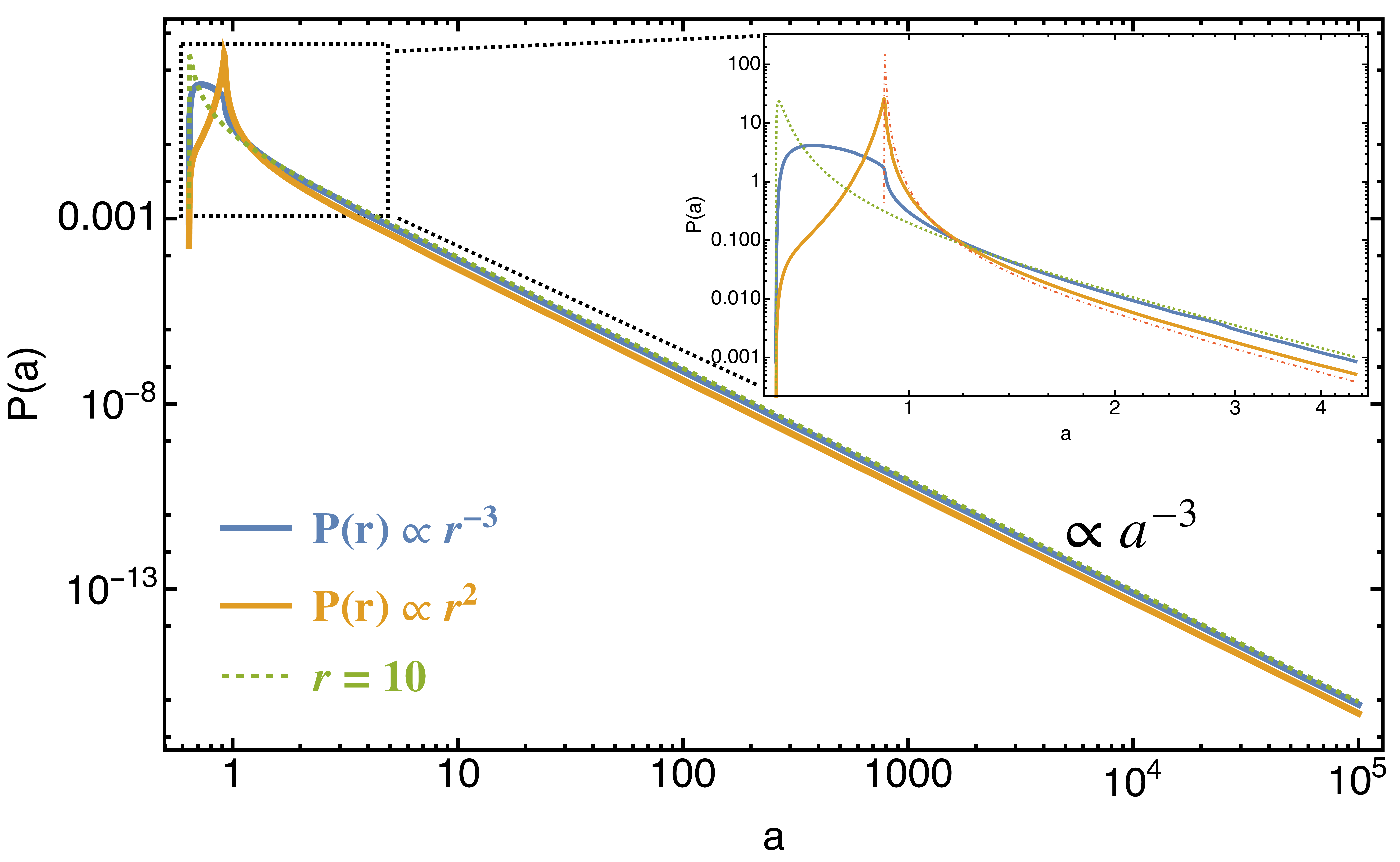}
\end{subfigure}
\begin{subfigure}{0.495\textwidth}
\centering
\vspace{0.3cm}
\includegraphics[width=1.01\linewidth, right]{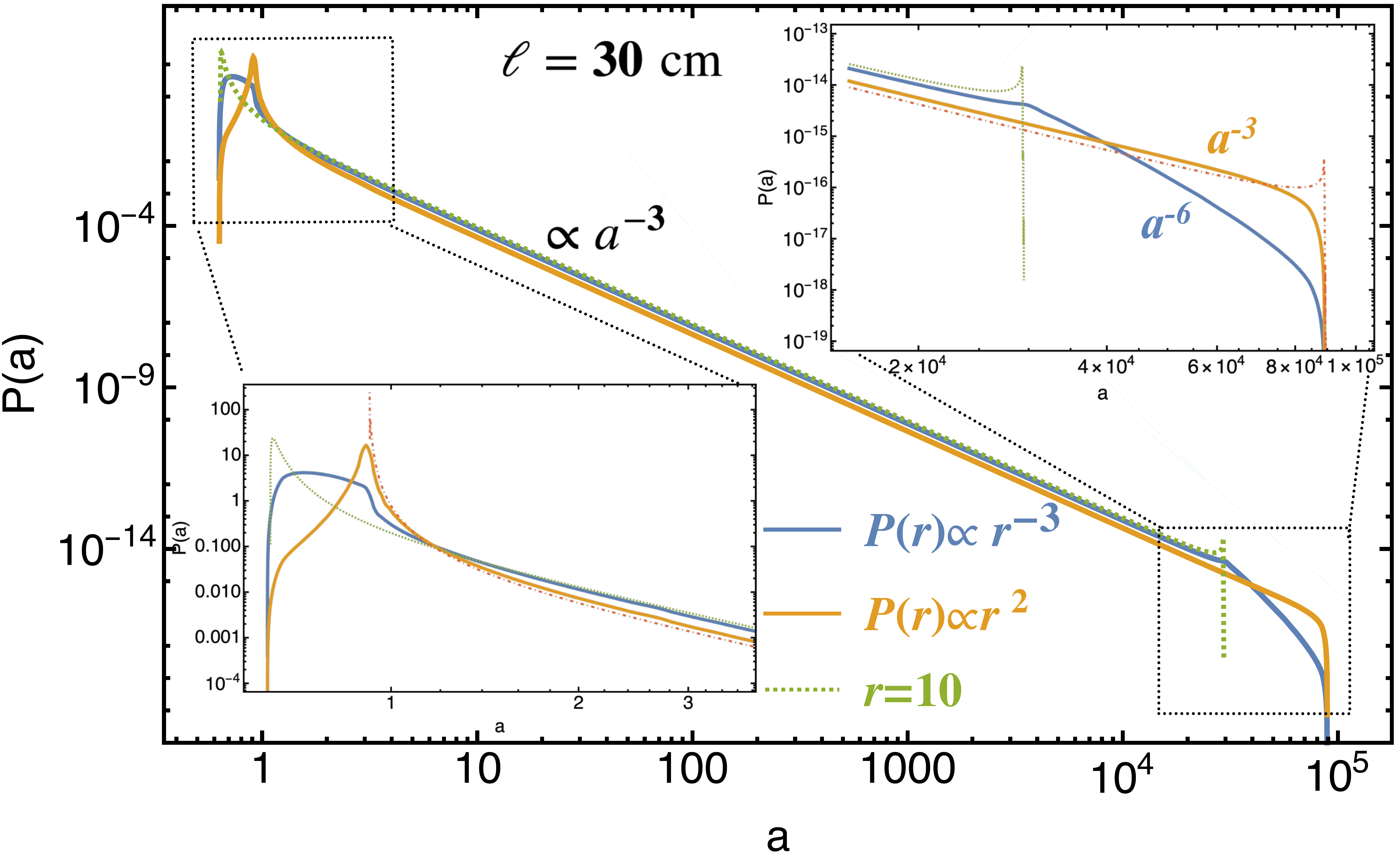}
\end{subfigure}
\caption{{\it Left}:~The amplification probability distribution, $P(a)$, for point-like sources in an active magnetosphere, in which events may occur at any~radius in the range $(10-50)R_g$, with probability $P(r) \propto r^{-k}$ (blue, $k=3$; orange, $k=-2$).~The $P(a)$~at a fixed $r=10$ (green, dashed) is shown for reference.~The radial distribution of the events affects mostly the low-amplification~peak, which receives contributions from individual peaks at different radii, with different relative weights (depending on $k$).~The~inset also shows the other extreme of $P(a)$ at $r=50$ (red dashed).~At large amplifications the $P(a) \propto a^{-3}$ scaling is unaffected for any $k$, besides a tiny decrease in the normalization.~{\it Right}:~same as left panel, but for finite-size sources ($\ell =30$~cm).~The effect at low-amplifications is identical (for the same $k$-value), but the source finite-size may leave an additional imprint: if $k>0$, the high-amplification end of the distribution steepens to $P(a) \propto a^{-(3+k)}$, beyond $a_{\rm max}(r=10)$ (shown by the green dotted curve).~The steeper power-law cuts off sharply at $a_{\rm max} (r=50) \approx 8.9 \times 10^4$ ($M/1.4M_{\odot}$)$^{1/2}/ (\ell/30~{\rm cm}$),~the maximum possible amplification.~If $k<0$, though, the slope in the high-amplification end remains unchanged all the way to $a_{\rm max} (r=50)$.} 
\label{fig:radial}
\end{figure*}
Moreover, we calculated the $P(a)$ resulting from allowing the hotspot, with fixed size $\ell$ and colatitude $\xi$, to~go~off~at any radius in~the range $r=10-50$~with~probability 
$P(r) \propto r^{-k}$. Results~are depicted in Fig.~\ref{fig:radial}, right panel, for two values of $k= -2, 3$.

~Contrary to point-like sources, the finite size $\ell$ introduces new features in the $P(a)$, due to the $r$-dependence of the maximum amplification, $a_{\rm max}(r) = 2 f(r)/r\theta_s(r)$.~First, the $P(a)$ spans a wider range extending to $a_{\rm max}(50)>a_{\rm max} (10)$. 
Second,~the 
 scaling $\propto a^{-3}$ is preserved only up to~$a_{\rm max}(10)$. Larger amplifications can only be achieved at larger radii, thus the corresponding section in the $P(a)$ is, in general, sensitive to the shape of~$P(r)$.~However, only $k>0$ determines a steepening to $P(a) \propto a^{-(3+k)}$ above $a_{\rm max}(10)$.~The $a^{-3}$ slope is preserved, instead, if $k<0$.~Eventually, a cut-off at $a_{\rm max} (r=50)$ appears, as larger radii - hence larger amplifications - are ruled out in our example.~Note that, since 
the amplification is mostly sensitive to $\theta_s=\ell/R$, the effect of a distribution in~$\ell$ will be (almost) degenerate with that in $r$:~however a small difference will remain, due to~the residual mild radial dependence of all coefficients in eqs.~(\ref{eq:aweaklens}) and (\ref{eq:afinitesize}).
\section{Probability of amplification for an active hotspot in the magnetosphere} 
\label{sec:rings}
The configuration discussed in this section is key to our scenario. It envisages
an active region, anchored to the magnetosphere, which as a result of~the NS rotation describes~an annulus of width $\ell$ centered on the spin axis. This, e.g., may be also consistent with the occurrence, in the Crab pulsar,~of GPs~always in the trailing edge of the main radio pulse, if GPs were  
interpreted as gravitationally-lensed `regular' pulses.~Possible implications for GPs will be discussed elsewhere. 
Here~we describe the probability of lensing in this configuration.

\begin{figure}[t]
\centering
\includegraphics[width=0.99\columnwidth]{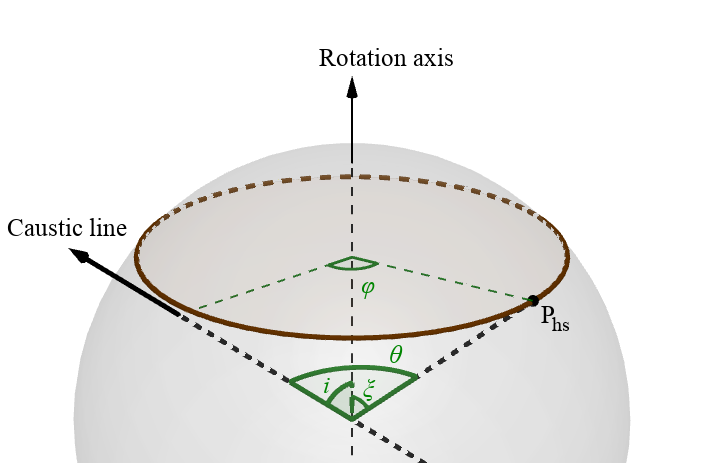}
\caption{Active hotspot geometry. The center of the hotspot, 
P$_{\rm hs}$, has angular distance $\xi$ from the rotation axis, with which the caustic line forms an angle $i$.~The observer lies at the opposite of the caustic line, relative to the NS.~As the NS rotates, the angular distance of~P$_{\rm hs}$ from the caustic goes from a minimal $\theta_{\rm min}$ (maximal amplification $a_{\rm max}$) at phase $\varphi = 0$, to a maximal $\theta_{\rm max}$ (minimal amplification~$a_{\rm min}$) at $\varphi = \pi$.~The rest of the rotation is symmetric (barring the slight asymmetry due to relativistic Doppler, which is negligible for NS spin periods longer than a millisecond; see \citealt{Bakala2023}).}
\label{fig:hotspotgeoangles}
\end{figure}
Consider an emission 
region
$P$ of finite size $\ell$, co-rotating with the star at radius $r$ and colatitude $\xi$ (Fig.~\ref{fig:hotspotgeoangles}). The 
probability of a given amplification will reflect the 
probability~of the corresponding separation angle, $\theta$, between~$P$ and the caustic line throughout the rotation,  %
\begin{equation}
\label{eq:theta}
   \cos{\theta} = \cos{\xi} \cos{i} + \sin{\xi} \sin{i} \cos{\varphi} \, .
\end{equation}
Here {\it i} is the angle between the rotation axis and the caustic, and $\varphi$ the rotation phase measured from~the~point~where~$P$~is closest to the caustic line.

The amplification $a$ now depends, through $\theta$, on $\varphi$, $i$ and~$\xi$, 
the latter two angles being fixed for a given FRB source.~A~key
\begin{figure*}[ht]
\centering
\begin{subfigure}{0.495\textwidth}
\centering
\includegraphics[height=0.61\linewidth, inner]{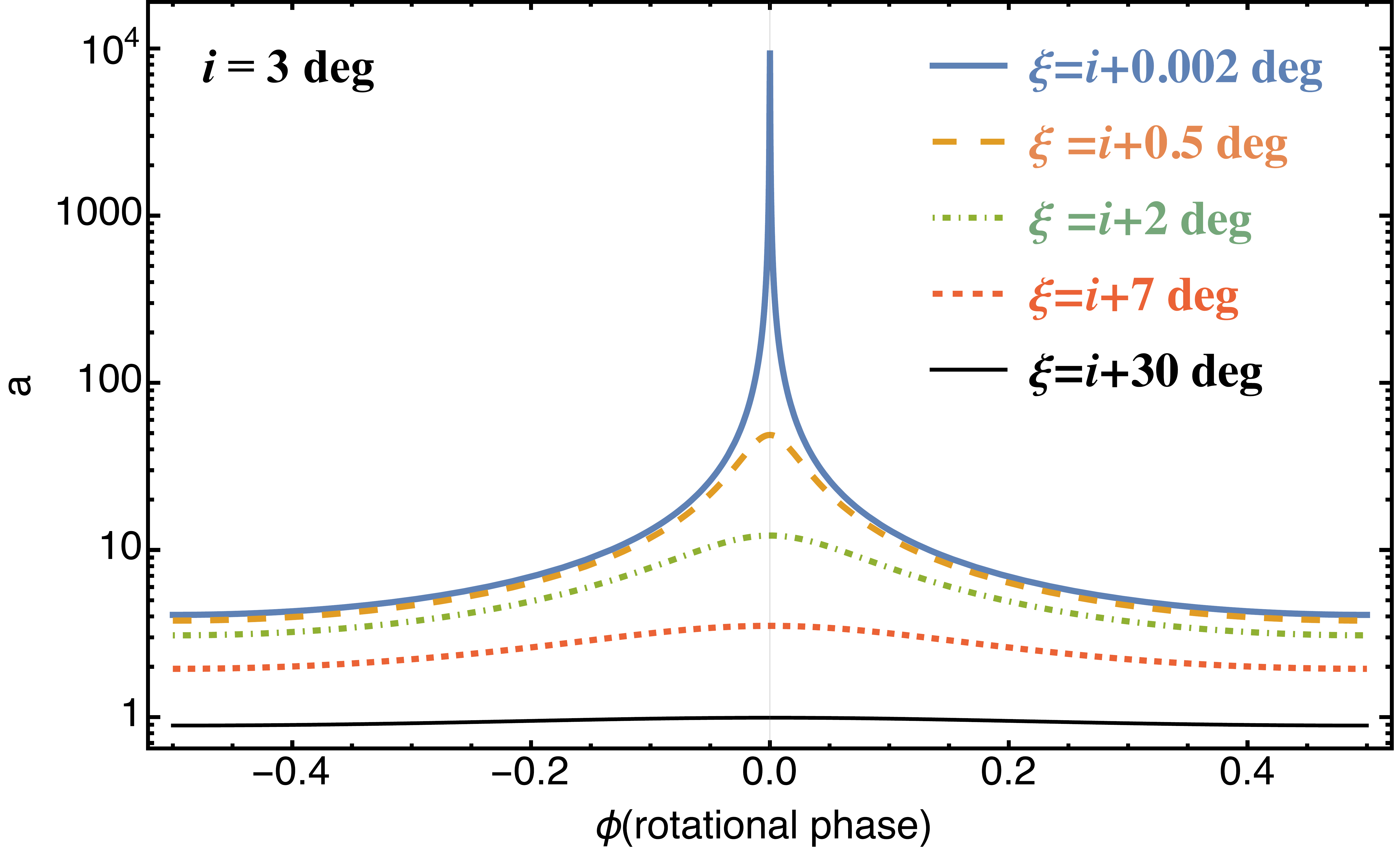}
    \end{subfigure}
\begin{subfigure}{0.495\textwidth}
\centering
\includegraphics[height=0.61\linewidth, right]{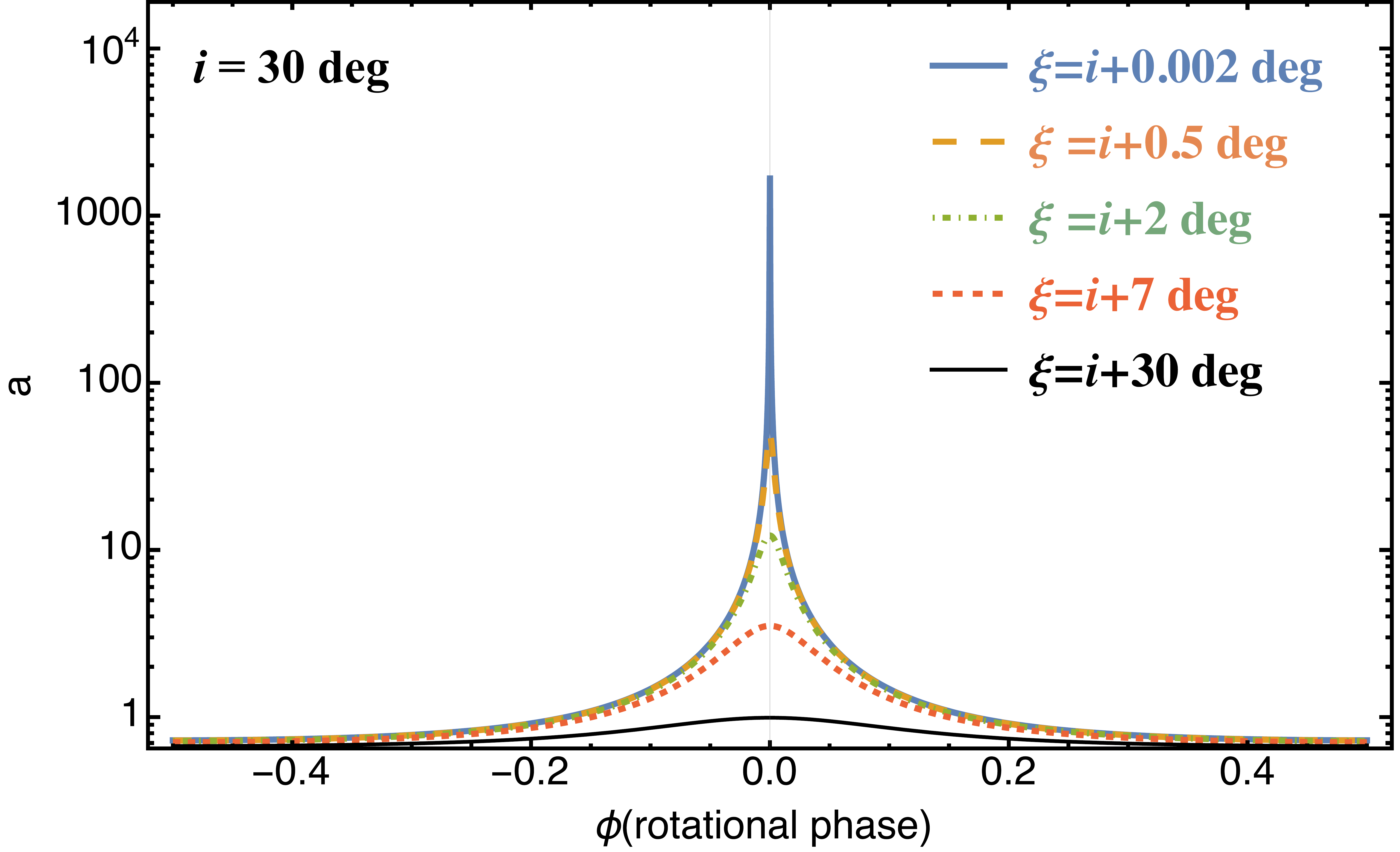}
    \end{subfigure}
\caption{
Examples of $a$-modulations vs. rotational phase, $\varphi$, for different source geometries.~\textit{Left Panel:}~a small inclination ($i = 3$ deg), and a co-latitude of the emitting region $\xi = 3.002, 3.5, 5, 10$ deg  
for the solid, dashed, dot-dashed and short-dashed curves, respectively.~\textit{Right panel:}~a larger inclination ($i=30$ deg) and the~same $|\xi - i|$ as in the left panel.~The~maximum amplification is the same in both panels.~However, the amplification drops quickly below unity at large $i$, while at low~inclinations,~it is relatively high over a wide $\varphi$-range.~The lower, solid black curves  
are representatative of more common configurations in a random sample, i.e. those a large value of $|\xi -i|$. }
\label{fig:temporale}
\end{figure*}
point is that there exist some (rare) combinations of $i$ and $\xi$, for which $\theta$ is small throughout one rotation, and the amplification correspondingly large, producing sources with frequent and highly-amplified bursts.~Conversely,  
amplifications will rarely occur if $\theta$ is large most of the time, resulting~in~very sporadic or altogether unobservable bursts.~It is precisely this dependence on the geometrical configuration that delineates the difference between repeating and non-repeating sources,~as discussed extensively in sec.~\ref{sec:repeaters}.~Fig.~\ref{fig:temporale} illustrates the variation of $a$ with the rotational phase, for different source orientations, when $r=10$.~The left panel shows variations of $a$ with $\varphi$ for a fixed (small)~inclination of the source, $i = 5$~deg,~and various co-latitudes of the emitting spot.~The right panel shows the same variation for a more inclined source ($i=30$ deg).%

As implied by eq.~(\ref{eq:theta}), $i+\xi$~sets the maximum distance from the caustic line, thus a large $i$ (or $\xi$) will suffice to~keep the minimum $a$ close to, or below, unity.~On the other hand, very large amplifications are only achieved when $i$ and $\xi$ are extremely close, regardless of their values.~However, when both angles are small, the high-amplification peak in the phase profile is broad, while it gets very narrow and sharp if both angles are large (and approximately equal).~In NS where both $i$ and $\xi$ are small, the emission can reach high amplifications more frequently, and over a wider phase interval;~thus, for~a given detection threshold, finite-duration events will on average remain detectable for a longer time span.~Conversely, when the peak is narrow, large amplifications occur rarely and, even when they do, they only last for a tiny phase interval.~This is fully consistent with the bursts of repeating sources in the CHIME catalogue, which show longer durations than the rest of the population \citep{Pleunis2021}.
\subsection[General properties of P(a)]{General properties of $P(a)$} \label{ssec:ringsgeneric}
~Equation~(\ref{eq:theta}) allows to define the probability of a particular amplification, $a$, as a function of $i$ and $\xi$

\begin{figure*}[ht]
\centering
\begin{subfigure}{0.48\textwidth}
\centering
\includegraphics[height=0.62\linewidth, inner]{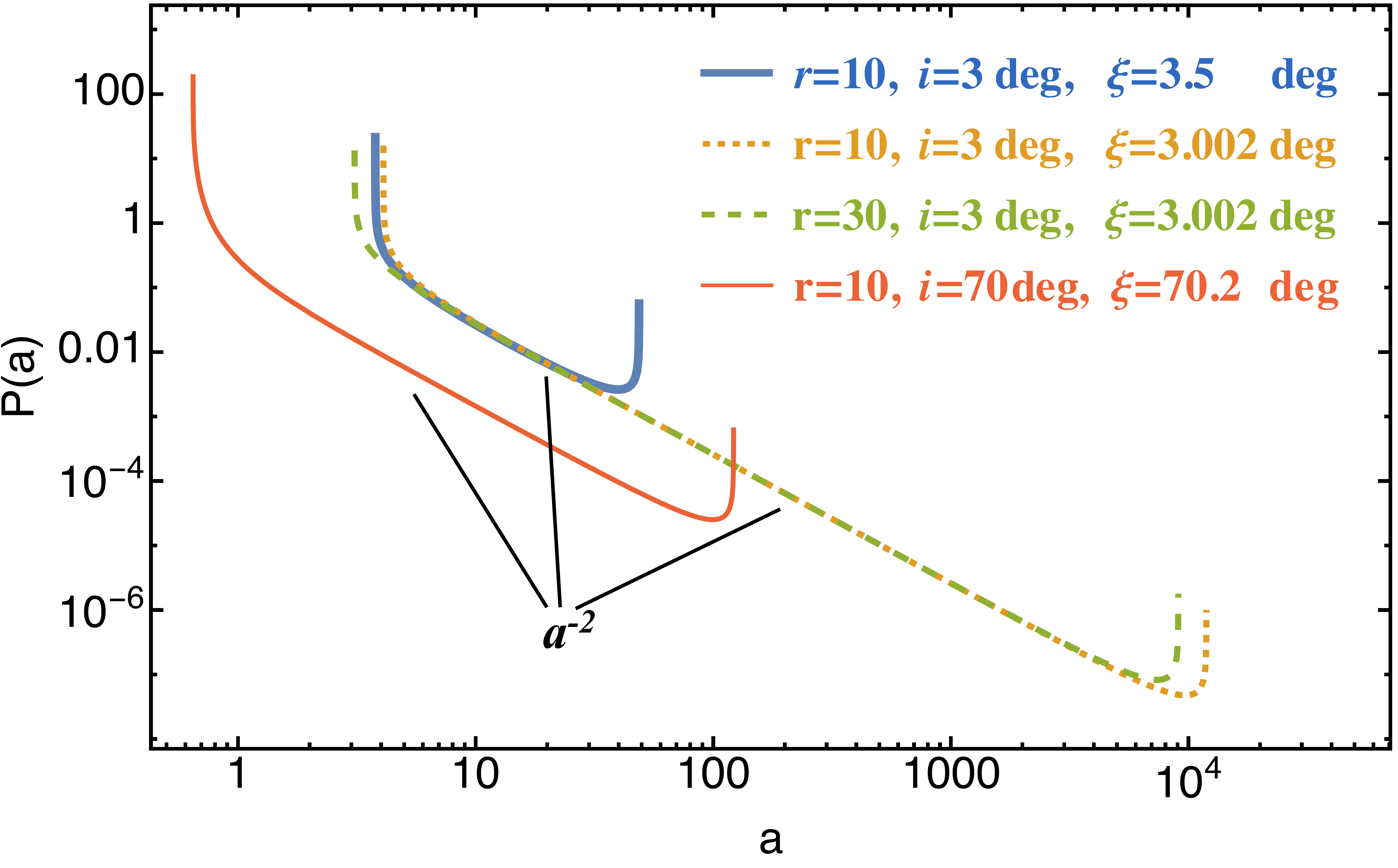}
    \end{subfigure}
\begin{subfigure}{0.48\textwidth}
\centering
\includegraphics[height=0.615\linewidth, right]{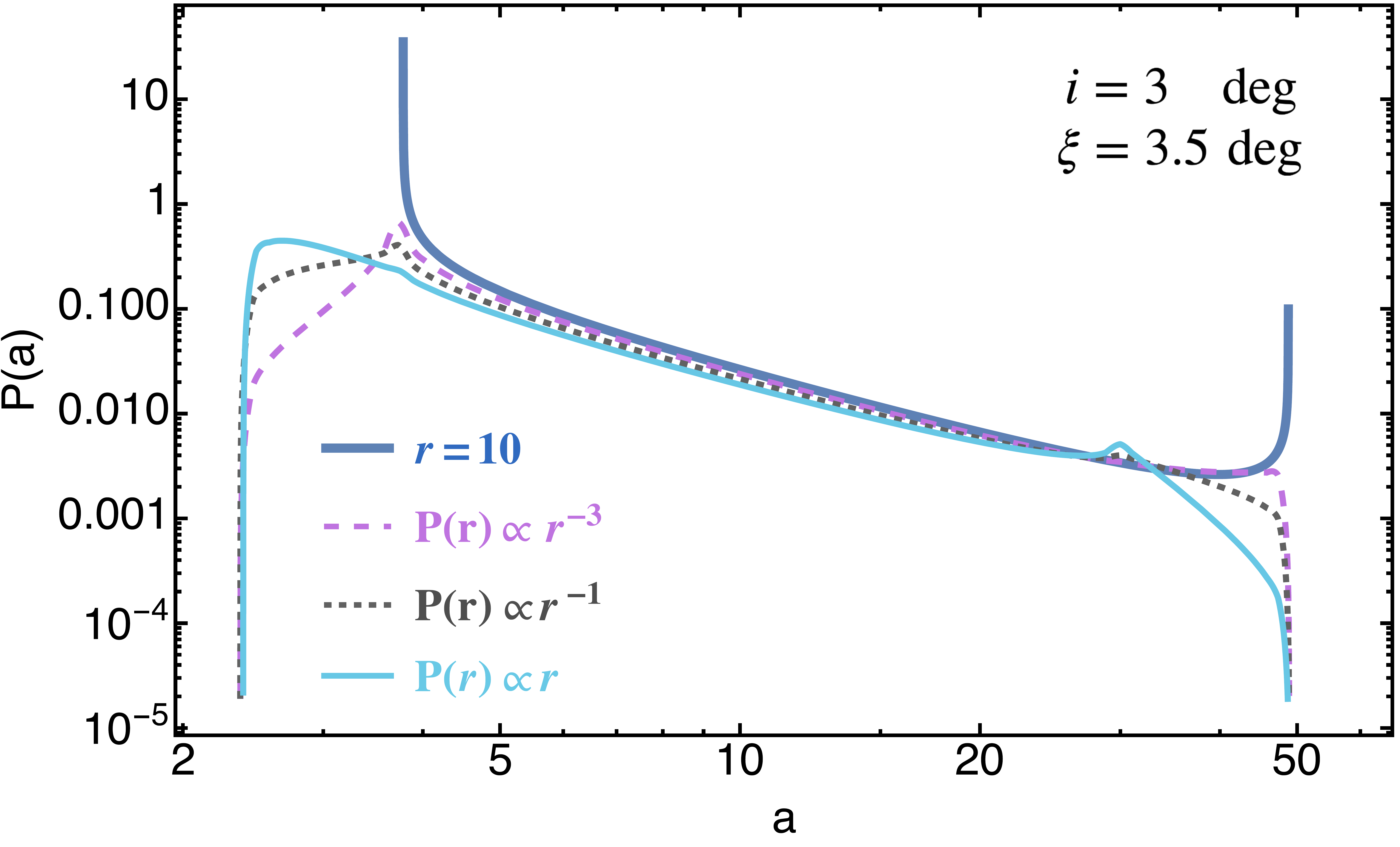}
    \end{subfigure}
\caption{{\it Left panel}:~Amplification probability, $P(a)$, for various configurations with $\ell=30$~cm.~{\it Blue solid:}~$r=10$,~$i=3$ deg, $\xi =3.5$~deg;~{\it Orange,~short-dashed:} $r=10$, $i = 3$ deg, $\xi =3.002$ deg.~{\it Green, long-dashed:} $r=30~$, $i = 3$ deg, $\xi =3.002$~deg;~{\it Red,~solid, thin:} $r=10$, $i = 70$~deg, $\xi =70.2$ deg.~{\it Right panel}:~effect on the original $P(a)$, i.e.~the solid blue curve from the left panel, of the hot-spot spanning a range of radii at fixed co-latitude~$\xi$.~At~low amplifications, the sharp edge of the blue curve is replaced by a much smaller peak and a smooth decay towards the minimum $a$, which becomes lower due to the decrease of $a$ with growing $r$, at a fixed $\theta$.~At large amplifications a steepening occurs, due to the same dependence, followed~by a sharp cut-off at the (unchanged) $a_{\rm max}$.~The three modified curves are for three probability distributions of the emission radius, indicated in the plot legend.}
\label{fig:pdiarings}
\end{figure*}
\be
\label{eq:system}
\begin{cases}
dl = R \sin \xi d \varphi \\
\sin \theta  d \theta = \sin i \sin \xi \sin \varphi d \varphi \\
P\left(\theta \right) d\theta = 
\displaystyle \frac{dl\left[\varphi(\theta)\right]}{\pi R \sin \xi} =\displaystyle \frac{d\varphi}{d\theta} \displaystyle \frac{d\theta}{\pi} \, , 
\end{cases}
\ee
where $dl$ is the arc length spanned by the emitting region for~a change $d\varphi$ in rotation phase.~The last of eqs. (\ref{eq:system}) leads to 
\be
P(a) =  \frac{1}{\pi}\displaystyle \frac{d\varphi}{d\theta}
\displaystyle \frac{d\theta}{da} \, ,
\ee
where we used the dependence $\theta(a)$ and restricted $\varphi$ to the interval $(0, \pi)$ where $\theta$ is uniquely determined (see Fig.~\ref{fig:hotspotgeoangles}).%

From Eq.~(\ref{eq:theta}) we get $d\theta/d\varphi = 
\sin i \sin \xi \sin \varphi/\sin \theta$,~hence 
\be
\label{eq:pdia-intrinsic}
P(a)da = \displaystyle \frac{\left(d\theta/da\right)
\sin \theta}{ \sin i \sin \xi \sin \varphi} \displaystyle \frac{da}{\pi}  \, ,
\ee
again determined by the relation $a(\theta)$.~Through a  
full rotation $\theta$ goes from $\left|i-\xi \right|$ to $i+\xi$ (Eq. \ref{eq:theta}); the emission annulus is~thus entirely in the ELR in all configurations with $i+\xi <\theta_{\rm we}(r)$.~In 
all other configurations, parts of the emission annulus will be 
in the weak lensing regime, and the corresponding~$d\theta/da$ must be obtained through eq.~(\ref{eq:aweaklens}). 

The left panel of Fig.~\ref{fig:pdiarings} shows our results for four selected configurations and a source size $\ell = 30$~cm.~Three of these~are entirely in the ELR ($i+\xi < 10$ deg): their $P(a)$'s have (i)~an 
extended power-law $\propto a^{-2}$, as opposed to the $a^{-3}$ scaling~of the previous section, 
due to the reduced dimensionality of 
the locus of the hot-spot potential positions relative to the caustic 
(a 
ring, as opposed to a spherical surface)
and~(ii)~two sharp rising edges 
at~the minimum and maximum amplification of each configuration.~Such edges are due to the dependence of 
$P(a)$ 
on $d\varphi /d \theta$, which steepens sharply at the extrema, where $\sin \varphi \approx 0$.~The orange,~solid~curve~shows a more generic configuration, with $\theta$ varying from 0.2~deg~(ELR)~to 140.2 deg (WLR); correspondingly, the low-amplification end rises more gradually towards the edge.

The right panel illustrates the result of allowing the hot-spot to span a range of radii ($r=10-50$) with $P(r) \propto r^{-k}$ and three different values of $k=-2, 1$ and 3.~In this case, extending the radial range does not, in general, allow the system to reach larger amplifications since the maximum amplification depends on $\theta_{\rm min}$, and the latter is set by $|i-\xi|$ alone, as long as 
this exceeds the hot-spot angular size, i.e. 
$ | i - \xi | > \theta_s \approx 0.03~{\rm deg}~ (\ell/10{\rm m}) (r/10)^{-1}$.~Only for smaller values of $|i-\xi|$~can the changing radius affect $a_{\rm max}$.~However, 
since the amplification depends on $r$ at fixed $\theta$, albeit weakly, the minimum amplification is somewhat reduced.~As
a result, both sharp edges of the distribution obtained with $r$ fixed get now spread over~a range of $a$-values.~In particular, the low-$a$ end of the distribution now contains a larger probability, at~the expense of~the high-amplification~end:~this happens because lower amplifications are increasingly more likely at larger radii.~To avoid adding more  degrees of freedom, we will keep $r$ fixed at $10$ in the following.%

\subsection{Latitudinal width of the active region} \label{app:ringwidth}
\begin{figure*}[ht]
\centering
\begin{subfigure}{0.48\textwidth}
\centering
\includegraphics[height=0.62\linewidth, inner]{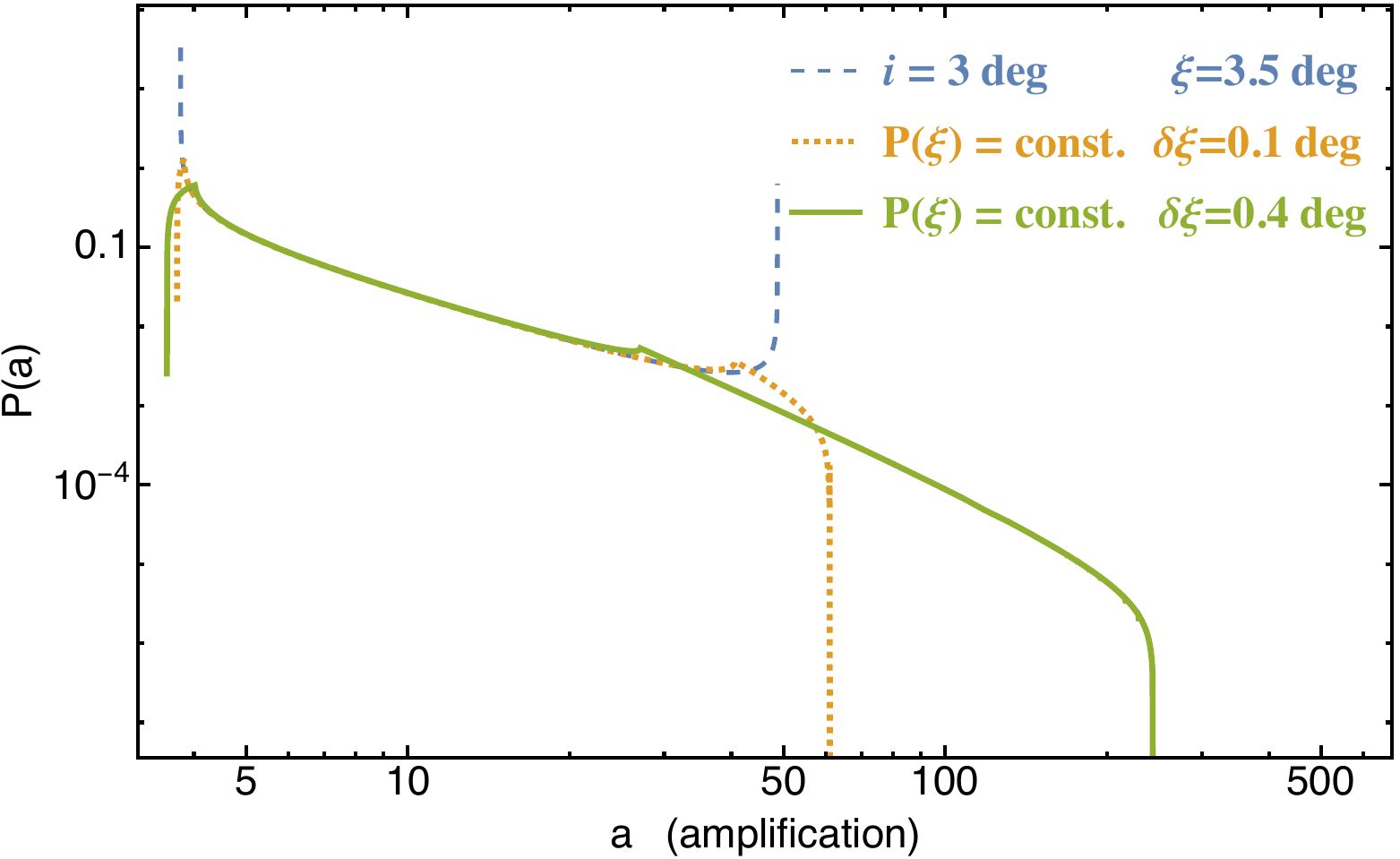}
    \end{subfigure}
\begin{subfigure}{0.48\textwidth}
\centering
\includegraphics[height=0.615\linewidth, right]{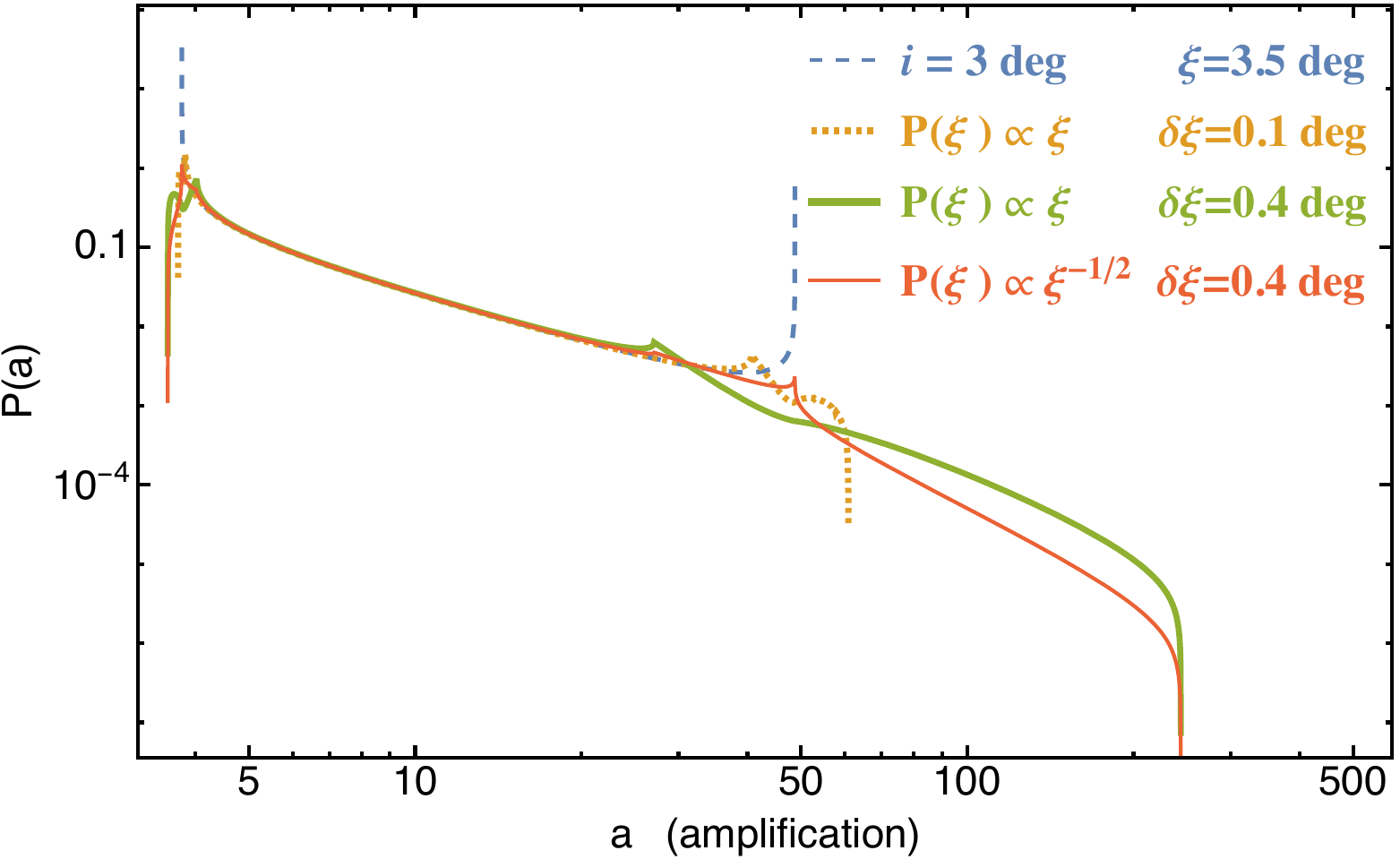}
    \end{subfigure}
\caption{Examples of hot-spot~jittering applied to the first configuration of Fig.~\ref{fig:pdiarings} (blue curve).~The modified~$P(a)$ reaches~a somewhat lower minimum and, at the same time, extends to larger amplifications, steepening towards the large amplification end. {\it Left panel}: a constant probability of jittering ($w=0$).~Curves are simply understood in terms of a transition from a thin~annulus, with $P(a) \propto a^{-2}$, to a wide annulus, $P(a) \propto a^{-3}$.~{\it Right panel}:~the shape of $P(\xi^\prime)$ leaves a specific imprint in the modified $P(a)$, causing slope variations in the high-amplification end of the distribution.~The slope may get flatter than $a^{-2}$ at intermediate points, or steeper than $a^{-3}$ around $a_*$ and close to the maximum amplification.}
\label{fig:jittering}
\end{figure*}

If the emitting spot is not static, but jitters randomly within an area of latitudinal extension $\delta \xi > \theta_s$, then it describes an annulus with an effective width $\delta \xi$. The resulting $P(a)$ is 

\be
\label{eq:jittering}
P(a; i, \xi, r) =  \int_{-\delta \xi}^{+\delta \xi} P(a; i, \xi+\xi^\prime, r) P(\xi^\prime)  d\xi^\prime \, ,
\ee
for which it is necessary to model the probability distribution associated to jittering, \textit{i.e.}\ $P(\xi^\prime)$.~To this aim,~we~parametrized $P(\xi^\prime) \propto \xi{^\prime}^{w}$ and tested three values, $w=-1/2,0,1$, applying eq.~\ref{eq:jittering} to one of the configurations of the previous~section, which has $i=3$ deg and $\xi=3.5$ deg, hence $\theta_{\rm min} =0.5$~deg (blue curve in the left panel of Fig.~\ref{fig:pdiarings}).~Moreover, for each $w$-value, 
we tried two possible widths, $\delta \xi=0.1$~deg or $\delta \xi = 0.4$ deg, for the annulus.~Results illustrating how the original $P(a)$  (blue, long-dashed curve) gets modified by each of the six combinations of $w$- and $\delta \xi$-values are reported in the two panels of Fig.~\ref{fig:jittering} (see plots and caption for details).

The
$P(a)$ of the ring gets now stretched~at~both ends, due~to jittering taking place both outwards or inwards relative~to~the original latitude $\xi$.~The case of a constant probability of jittering ($w=0$) is depicted in the left panel:~we note an interesting 
transition of the $P(a)$~from~$\propto~a^{-2}$ to~the steeper $\propto a^{-3}$ scaling, which occurs at a characteristic amplification, $a_*$, determined~by the width of the annulus.~This transition can be understood as follows:~as long as $\delta \xi  \ll \theta_{\rm min}$, the annulus is ``thin'' and $P(a)$ maintains the characteristic $a^{-2}$ scaling of ring-like structures 
in the ELR.~As $\delta \xi > \theta_{\rm min}$, on the other hand, the annulus becomes ``wide'' and increasingly akin to a two-dimensional surface.~The~transition to the $a^{-3}$ scaling thus occurs at $a_* \approx a(\theta_{\rm min}+\delta \xi, r)$.~In the left panel of Fig.~\ref{fig:jittering}, the transition for the thin annulus ($\delta \xi=0.1$ deg) 
is at $a_* \lesssim 40 $, and is closely followed by a cut-off since $\theta_{\rm min}$ only decreases from 0.5~to~0.4~deg.~For the wider annulus, on the other hand, we have a lower $a_* \approx 25$, and the $a^{-3}$ slope extends much further before cutting off at $a_{\rm max} (\theta=0.1{\rm deg})\approx 250$.

Additional striking features appear at both extremes of the $P(a)$ when the probability of jittering is a function of $\xi$ (right panel in Fig.~\ref{fig:jittering}).~Focusing on the high amplification end, we observe an early drop steeper than $a^{-3}$ around $a_*$, followed by a flattening of the profile up to $a_{\rm max} (\theta = \xi-i-\delta \xi)$, at which the $P(a)$ eventually cuts off to zero.~A wider range $\delta \xi$ for the jittering makes this feature more apparent and extended.~These properties are likely to play a role in detailed models of the energy distribution in individual sources \cite[see, e.g., results on FRB 20201124A presented by][]{Kirsten2024}.~A thorough investigation will be carried out in a future study.~In the rest of this paper, however, we restrict our attention to the case of a fixed $\xi$, to avoid introducing additional new assumptions to our calculations.

\subsection[A population of randomly-oriented sources]{A population of randomly-oriented sources: \^{P}(a)}
\label{sec:cappello}
We have shown that the amplification probability for an~individual source in~the ring-like geometry is $P(a) \propto a^{-2}$.~What happens, though, to the probability of amplifications in a large population of sources, each with its own random orientation ($i$) and random hot-spot co-latitude ($\xi$)?~Within such a population,
\begin{figure}[t]
\centering
\includegraphics[height=0.645 \linewidth, inner]{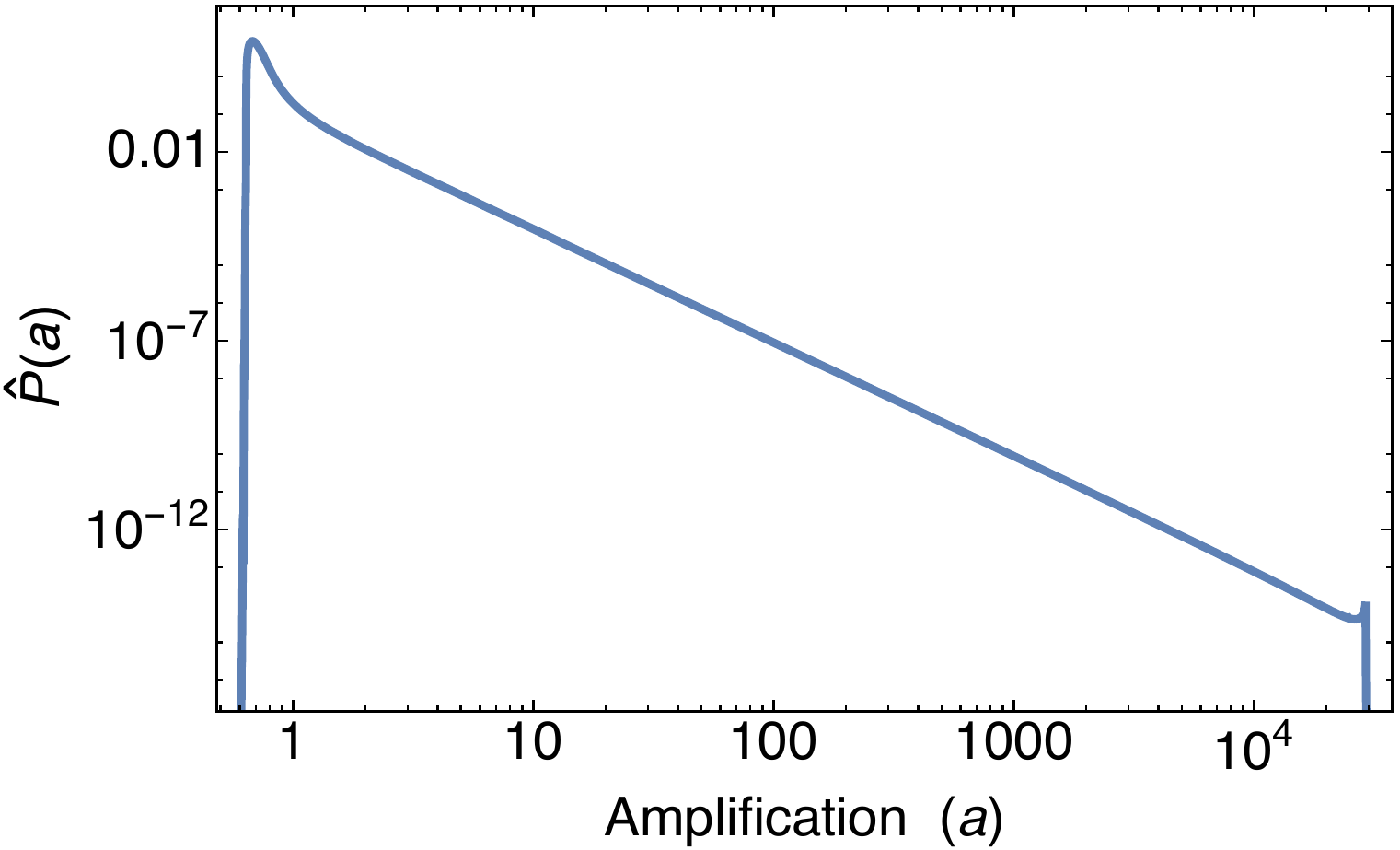}
\caption{Collective probability distribution of amplifications, $\hat{P} (a)$, in a population of randomly-oriented sources, each having a co-rotating hot-spot at random co-latitude~$\xi$.~For $a>2$ we have $\hat{P}(a) \propto a^{-3}$, even though in each individual source $P(a) \propto a^{-2}$.~This $\hat{P}(a)$ is almost identical to that of~a single source in which~the hot-spot can occur randomly within a spherically symmetric region of the magnetosphere.~This stems from the spherical symmetry introduced, within the population, by the random orientation of individual sources.} 
\label{fig:cappello}
\end{figure}
each $(i,\xi)$-configuration will occur with probability 1/4 $\sin i \sin \xi$.~Thus, multiplying the probability of a given $(i,\xi)$-configuration by its $P(a)$, and integrating over all possible configurations, we obtain $\hat{P}(a)$, the probability distribution of $a$ for the ensemble of randomly-oriented sources,
\be
\hat{P}(a) = \int_0^{\pi/2} \int_i^{\pi-i}  P(a; i, \xi)\sin i \sin \xi \,d\xi \, di\, ,
\ee
where we used the symmetries in the system to limit the integration range of each variable.~The resulting 
$\hat{P}(a)$, depicted~in Fig.~\ref{fig:cappello} (solid blue curve), scales like a power-law $\propto a^{-3}$, despite the individual $P(a)$'s being flatter ($\propto a^{-2}$)~in the adopted geometry.~This steepening, which is possible because the $P(a)$ of individual sources are truncated power-laws, can be understood as a result of larger amplifications being achieved only by sources with favourable combinations of $i$ and $\xi$, and the latter becoming increasingly more rare in the population.~The 
$\hat{P}(a)$ calculated here is almost~identical to the $P(a)$ of sec.~\ref{sec:magsphere} for a spherically-symmetric emission volume (right panel of Fig.~\ref{fig:regimes}):~effectively, the spherical symmetry of random $(i, \xi)$ combinations re-introduces, at the level of the population, the characteristic $a^{-3}$ scaling of spherically-symmetric sources.~As a result, in our model it is possible, or even expected, that the amplification distribution of events from an individual source differs from (be flatter than) that of the whole population.
\section{Observed vs. intrinsic properties}
\label{sec:obsvsint}
Knowing the form of $P(a)$ for various emission geometries, we can now calculate its imprint on~the luminosity distribution of lensed events.~In 
the simplest possible scenario we consider standard candle 
hot-spot 
events with intrinsic luminosity $\hat{L}_0$, hence 
a probability distribution $P_0 (L_0) \equiv \delta (L_0 - \hat{L}_0)$ for their intrinsic luminosities.~The 
probability distribution for amplified events with lensed luminosity $L$ will be
\begin{eqnarray}
\label{eq:PdL}
P(L) dL & = &\int_{L_{0,{\rm min}}}^{L_{0,{\rm max}}} P_0(L_0) dL_0 ~P\left[a\left(L\right)\right] \displaystyle \frac{da}{dL} dL = \nonumber \\
& = & \int \delta(L_0-\hat{L}_0) dL_0 ~P\left(\displaystyle \frac{L}{L_0}\right) \displaystyle \frac{dL}{L_0} = \nonumber \\
& = & \displaystyle \frac{P\left(L/\hat{L}_0\right)}{\hat{L}_0} dL = \displaystyle \frac{P\left[a\left(L\right)\right]}{\hat{L}_0}~dL \, ,
\end{eqnarray}
showing that the observed luminosity distribution will not track the intrinsic one but, rather, the probability distribution of the amplification factor.~The 
resulting $P(L)$ translates to~an observed ``luminosity function'' $F(L)$, {\it i.e.} the rate of~observed events at luminosity $L$, according to $F(L) = A P(L)$, where the normalization constant $A$ is determined by 
\be
\label{eq:normalise}
\int_{S_{\rm thr}}^{S_{\rm max}} F(S) dS = A \int_{S_{\rm thr}}^{S_{\rm max}} P(S) dS \equiv {\cal R}_b
\ee
where ${\cal R}_b$ is the rate of {\it observed} events above the detection threshold, $S_{\rm thr} = a_{\rm thr} \hat{S}_0$.~From the above it obtains that
\be
\label{eq:defineFdL}
F(S) = {\cal R}_b \displaystyle \frac{P(S)}{\displaystyle \int_{S_{\rm thr}}^{S_{\rm max}} P(S) dS}  =\displaystyle \frac{{\cal R}_b}{\hat{S}_0}~ \displaystyle \frac{P\left(S / \hat{S}_0\right)}{\displaystyle \int_{a_{\rm thr}}^{a_{\rm max}} P(a) da }\, .
\ee

In a more general case in which events have a range of
intrinsic luminosities, the probability of observing one with luminosity between $L$ and $L+dL$ is still given by eq. (\ref{eq:PdL}).~We do not specify the functional form of $P_0(L_0)$, but only assume that it extends from 
a minimum intrinsic luminosity $L_{0, {\rm min}}$ to a maximum $L_{0, {\rm max}}$.~Thus 
\be
\label{eq:PdLbis}
P(L) dL = \int_{L_{0, {\rm l}}}^{L_{0, {\rm h}}} P_0(L_0) dL_0 P(a)\frac{da}{dL} dL  \, ,
\ee
where the integral extends over all $L_0$-values that can achieve $L$ if amplified.~Hence, $L_{0, {\rm l}}~=~{\rm max} \left[L/a_{\rm max}, L_{0, {\rm min}}\right]$ and $L_{0, {\rm h}} = {\rm min} \left[L/a_{\rm min}, L_{0, {\rm max}}\right]$. Again, the function $P(L)$ must be normalized to the observed event rate (Eq.~\ref{eq:normalise}) in order to obtain the corresponding luminosity function, $F(S)$. 
Further details and discussion are presented in Appendix~\ref{app:pdevpda}.

\section{Repeating and non-repeating FRBs from a single source population}
\label{sec:implications}
In order to highlight the role of gravitational lensing~in~shaping the appearence of FRBs, we first
discuss in an approximate fashion some of the key constraints imposed by observations.~
In~our scenario,~the {\it observed} event rate per source, ${\cal R}_b$, is~lower than 
the intrinsic rate, ${\cal R}$, by the probability of amplification above the detection threshold, i.e. a factor $P(a>a_{\rm thr})$.~If the emission~is~also beamed by a factor $f_b <1$, then the two event rates~are~related as\footnote{See Appendix \ref{app:lumfunc} for further details.} ${\cal R}_b = f_b {\cal R} P(a >a_{\rm thr}) $.~Accordingly, the number of~{\it observed} events from a source ${\cal N}_b = {\cal R}_b T_{\rm obs}$, in the observing time $T_{\rm obs}$, will be smaller than the {\it intrinsic}~number~of events~emitted and beamed towards us, $N_{\rm int}$ $= {\cal R} f_b T_{\rm obs}$.

We constrain ${\cal R}_b$ following  \citet{cordes16}, who estimate a birth rate $\sim 10^4$~day$^{-1}$ for the cosmic~NS population within the Hubble volume $V_H = 4/3 \pi (c/H_0)^3$, when scaled from the galactic value.~As this is 
$\sim 10$ times 
higher than the all-sky FRB rate, $\Gamma_{b, {\rm obs}} \sim 500$ day$^{-1}$ (\citealt{chime20}), this population is evidently sufficient to  account for $\Gamma_{b, {\rm obs}}$, e.g. requiring that $\sim$ 5-10\% of all NS within $V_H$ produce one {\it observable} FRB in their active lifetime, $T_b$.~Because magnetars are estimated to represent~$\lesssim$~10\% of all~NS formed in core-collapse (e.g.~\citealt{gaensler05b, beniamini19}) let us suppose, for the sake of the argument, that all FRBs are due to magnetars, each leading on average~to~1~{\it observed} event in a time $T_b \sim 10^4$~yr, the characteristic age for~known active magnetars.~The average 
observed rate per source within the Hubble volume is thus
\be
\label{eq:estimate-rate}
\langle {\cal R}_b \rangle 
\approx 10^{-4}~{\rm yr}^{-1} ~\displaystyle \frac{\left(\Gamma_{b, {\rm obs}}/10^3~{\rm day}^{-1}\right)}{\left(T_b/10^4~{\rm yr}\right)} \, .
\ee
This is much lower than~the $\sim 30$ hr$^{-1}$ (2.6 $\times 10^5$ yr$^{-1}$)~rate 
of the most active repeaters 
currently known, 
FRB 202121102A and FRB~20201124A (averaged over active and `off' states). This huge rate mismatch represents a severe constraint for any model aimed at explaining FRB observations.~Taken at face value, it implies that~most sources may never be observed to repeat, and that just 
a handful 
 of the most active repeaters may exist~in the observable universe, in order not to exceed the all-sky rate, $\Gamma_{b, {\rm obs}}$.
 \begin{figure*}[t]
\centering
\includegraphics[height=0.81\columnwidth]{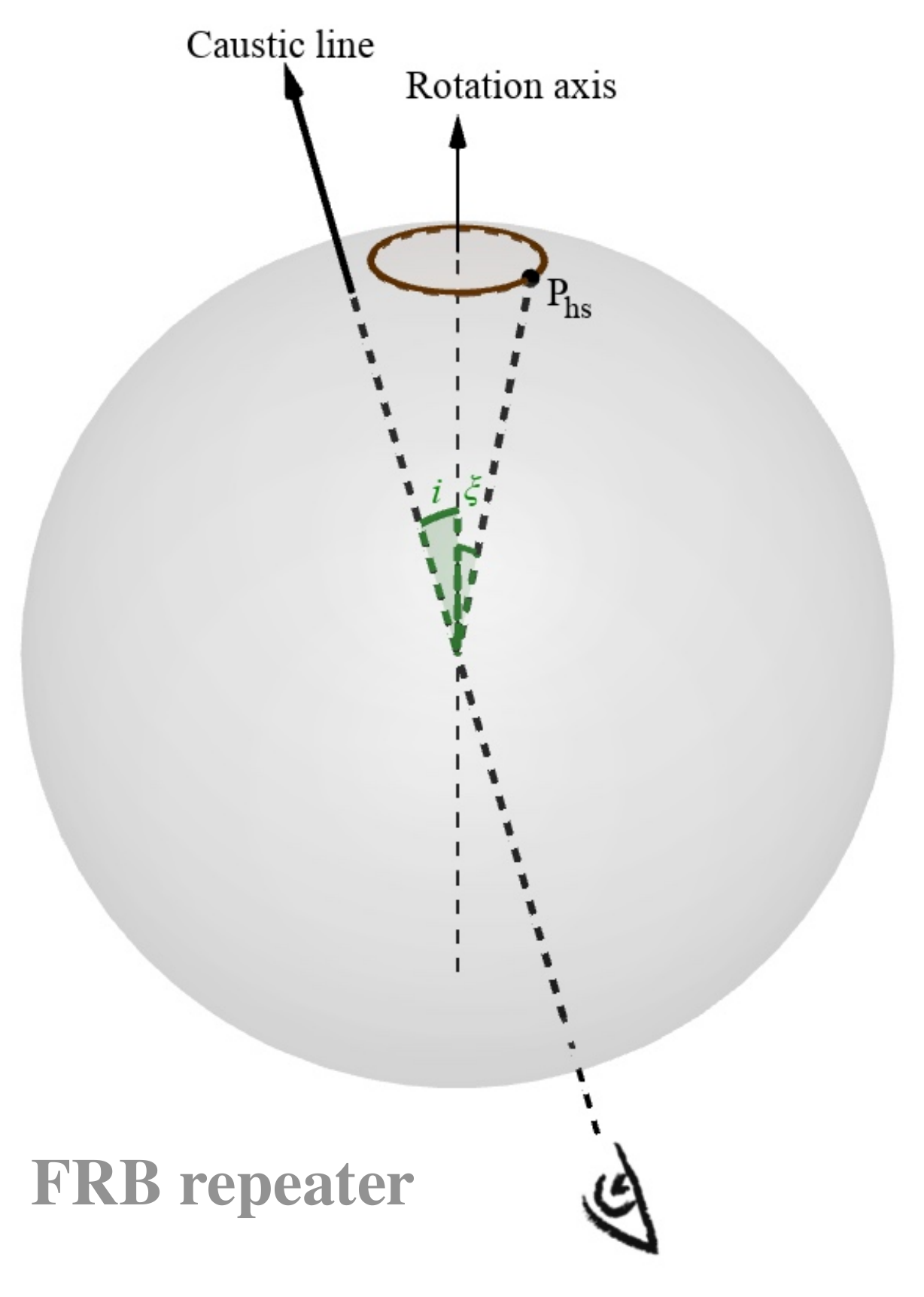}
\includegraphics[height=0.81\columnwidth]{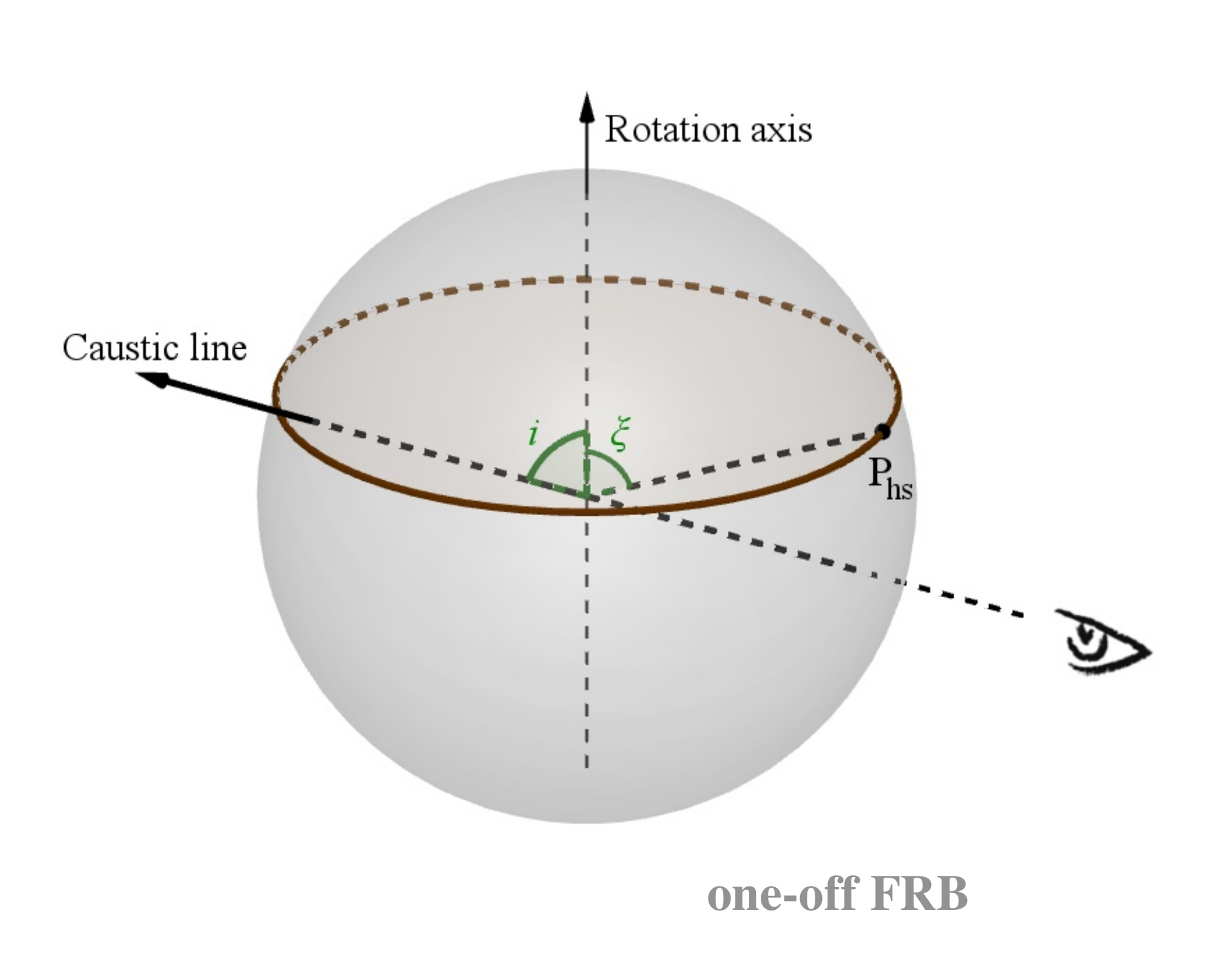}
\caption{Two different orientations of identical sources,~corresponding to an active repeater (left) and to a one-off source (right).~In the left panel, the hot-spot is located at a small angular distance from the rotation axis, $\xi \lesssim$ a few degrees, nearly identical to the misalignement of the line-of-sight~($i$):~the hot-spot stays close to the caustic line during most, or even all, of its rotation, hence a majority of its flares are sufficiently amplified to be detectable.~The right panel depicts a more common geometry and orientation, in which both $i$ and $\xi$ are relatively large and comparable to each other:~the hot-spot spends most of the time far from the caustic line, and only rarely its flares get amplified - hence detected.~Even more common, and more numerous, are orientations for which no amplification and no detection are ever achieved.}
\label{fig:repeater-oneoff}
\end{figure*}

Such a dichotomy between repeaters and non-repeaters can be 
accounted for 
in our NS self-lensing scenario if the hot-spot is stably anchored in the magnetosphere, i.e.~as~in~the~annular 
geometry of sec.~\ref{sec:rings}.~In 
that case 
the probability~of exceeding a given amplification threshold, $P(a>a_{\rm thr})$, depends very strongly on the source orientation 
(instead, no such effect arises in the spherical geometry of sec.~\ref{sec:magsphere}).~Thus, sources with a favourable orientation,~i.e. a large probability of exceeding $a_{\rm thr}$, may~lead to a~much larger number of {\it observed} events than~less favourably oriented sources, even if they had the same intrinsic rate and produced the same total events, $N_{\rm int}$.~In this framework, one-off FRBs are typical randomly-oriented sources, with very low probability of achieving amplifications above threshold.~Thus, even in the simplest assumption of identical FRB sources (i.e.~same event rate and energy), the ratio between~$\langle {\cal R}_b \rangle$ in eq.~\ref{eq:estimate-rate} and the $\sim$ 30 hr$^{-1}$ events of active repeaters has~a clear interpretation:~it represents the characteristic probability with which we detect typical events from unfavourably oriented sources (one-off FRBs) when they get sufficiently amplified.~Therefore, we estimate $P(a>a_{\rm thr}) \sim 10^{-4}/(2.6 \times 10^5) \sim 4 \times 10^{-10}$ which,~as~can be read off Fig.~\ref{fig:cappello}, implies a characteristic threshold $a_{\rm thr} \sim 10^3$.~The most energetic FRB ever recorded had~$E_{\rm iso} \gtrsim 10^{41}$~erg, for which our argument implies a true~energy $\sim E_{\rm iso}/a_{\rm thr} \sim 10^{38}$~erg, and $(4 \times 10^{-10})^{-1} \sim (3 \times 10^9) $ identical events that were missed as they didn't achieve $a\geq~10^3$.~The~resulting energy budget is $E_{\rm TOT} \sim 2.5\times 10^{47}$~erg for the most energetic one-off FRB.~Formally, this represents a lower limit to the actual energy budget, as the efficiency of radio emission is expected to be $\ll 1$ (it was $\sim 10^{-5}$ in the FRB-like event from the galactic magnetar SGR 1935+215;~e.g. \citealt{Mereghetti2020}).~However, the above calculation implicitly assumes that the source maintains its extreme burst rate for $T_b = 10^4$ yrs.~Evolutionary effects may well shorten the duration of such an~intense bursting phase, in turn reducing the energy estimate.~For instance, if the burst rate were to remain constant for $\sim 300$ yr and then decrease like $t^{-1}$, 
while keeping the same energy per event, the radio energy budget would decrease by a factor 10 (more details in Appendix~\ref{app:C}).

Conversely, active repeaters are ideally-oriented sources with $P(a > a_{\rm thr}) \approx 1$, for which most bursts are detected. We know only a handful of very active repeaters while, adopting the cosmic star formation rate of \cite{madau14}, we estimate $\sim 10^7$ magnetars younger than $\sim 10^2$~yrs within $z\approx 0.193$, the distance and likely age of FRB 20121102A \citep{Metzger2017,Yang2017,Hilmarsson2021}.~Hence, if all FRB sources~are identical, the most active repeaters must be very unlikely occurrences ($\sim 1$ in $10^{6}$ sources), although their enhanced detectability likely boosts their incidence among observed~FRBs.~A more accurate assessment would involve the modelling of a population of non-identical sources with a range of event rates and energies:~our crude estimate, though, establishes the extreme rarity~of very active repeaters.~These may exist only for finely-tuned combinations of $i$ and $\xi$, in turn  requiring\footnote{A more realistic calculation
may slightly relax this constraint on $(i, \xi)$.} both angles to be small ($<$ 4 deg) and almost identical ($|\xi - i| < 0.05$~deg).~Note~that the energy budget of the very active FRB 20121102A is estimated to be\footnote{If most events of such active repeaters are indeed detected, and at least mildly amplified, their true energetics may be somewhat lower.} $\sim 3 \times 10^{47}$ erg for $T_b \sim 10^4$~yrs (e.g.~\citealt{li21}), close to the above estimate for the most powerful~one-off.~In this case too, evolutionary effects may significantly reduce the source energy budget.
\subsection{Formal definitions for repeaters and one-off sources}
 \label{sec:repeaters}
To keep our treatment simple we consider in following the case in which hotspots are standard candles.~Under this assumption 
we can~express 
via eqs.~(\ref{eq:theta}) and (\ref{eq:system})   
the probability that bursts in a randomly oriented source overcome a given detection threshold ({\it i.e.} achieve $a \ge a_{\rm thr}$) at least $1/N_{\rm int}$ of the times.  
\begin{figure*}
\centering
\begin{subfigure}{0.495\textwidth}
\centering
\includegraphics[height=0.62\linewidth, left]{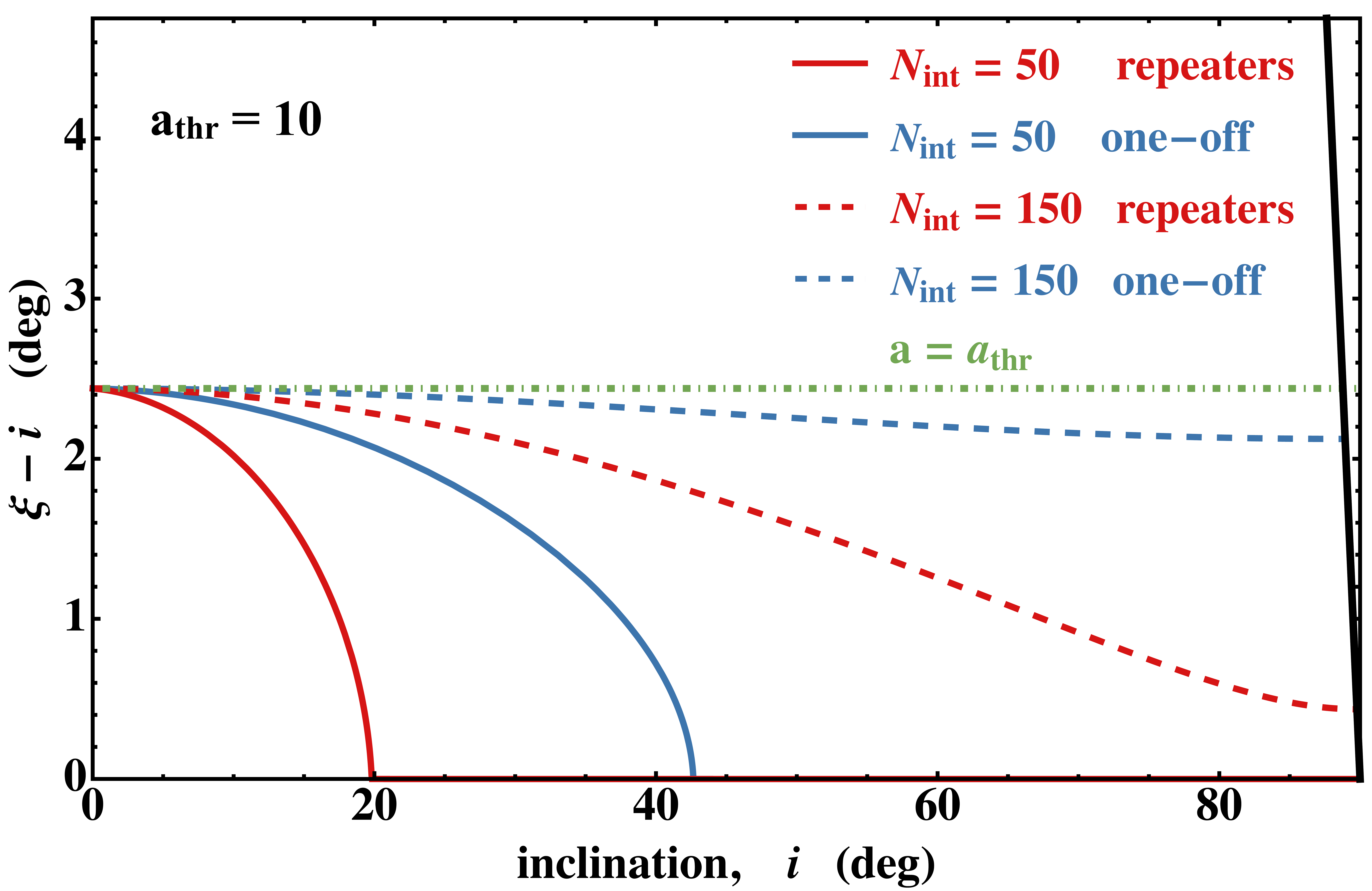}
\centering
\includegraphics[height=0.62\linewidth, left]{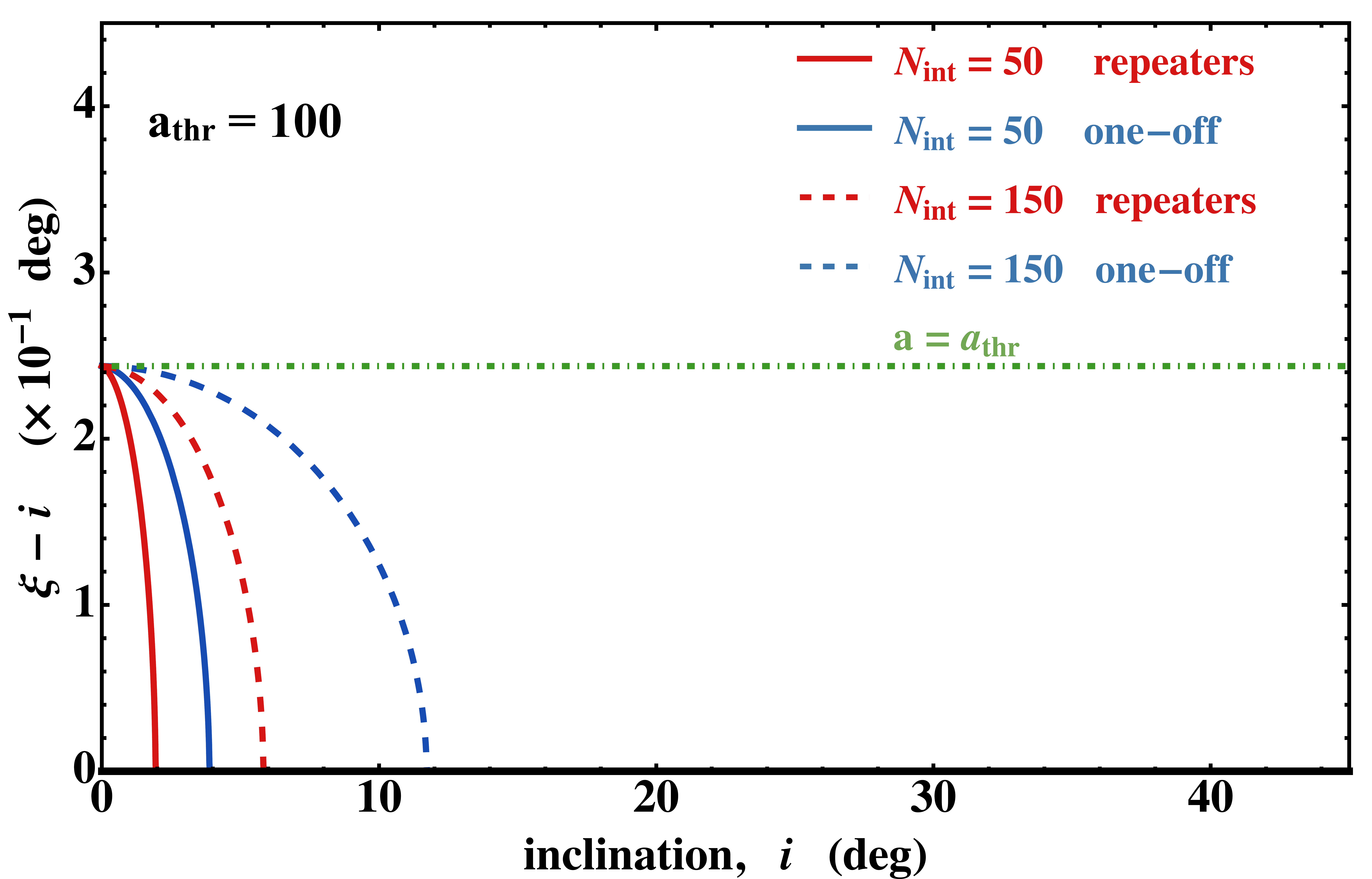}
\end{subfigure}
\begin{subfigure}{0.495\textwidth}
\centering
\includegraphics[height=0.62\linewidth, right]{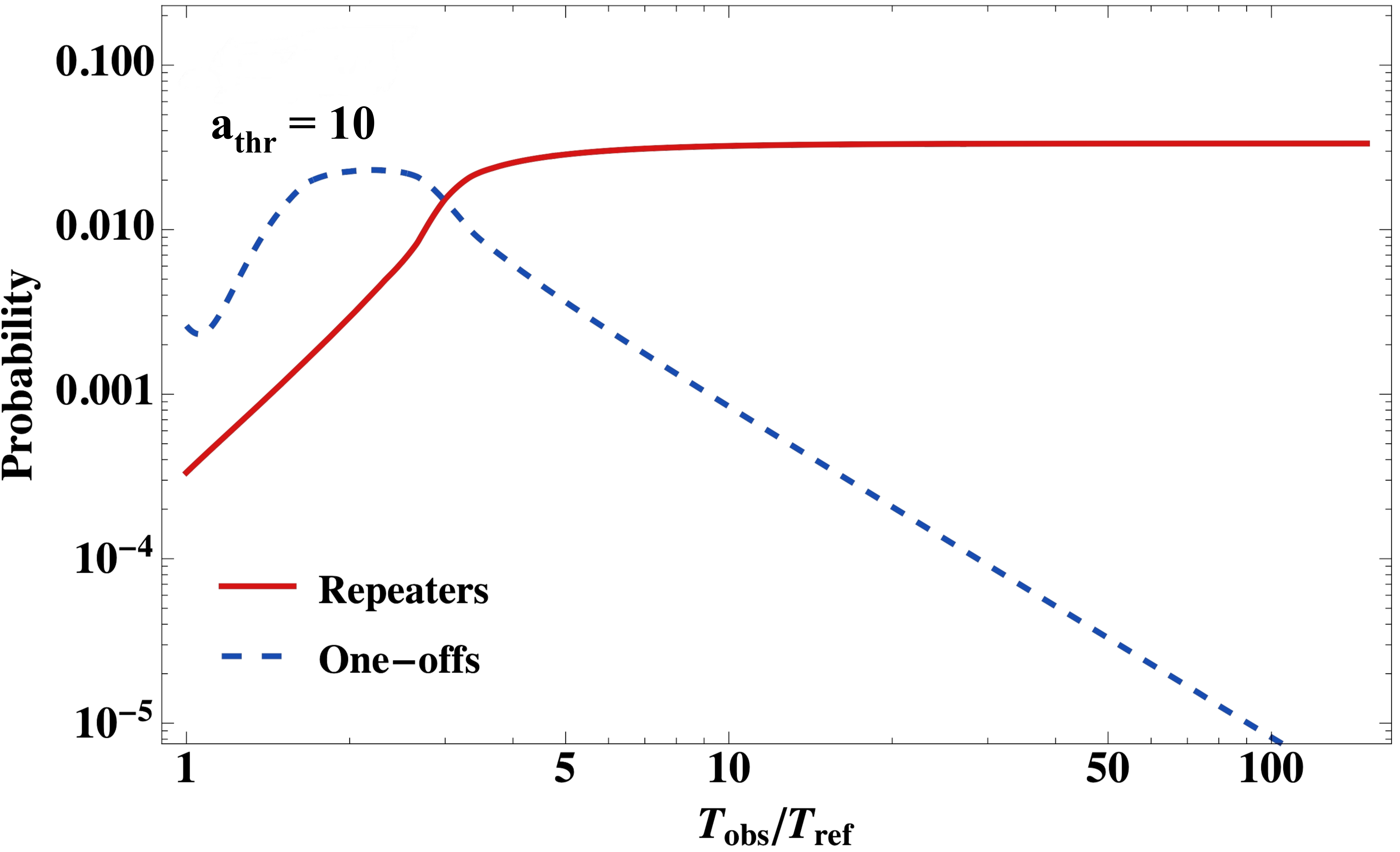}
\includegraphics[height=0.62\linewidth, right]{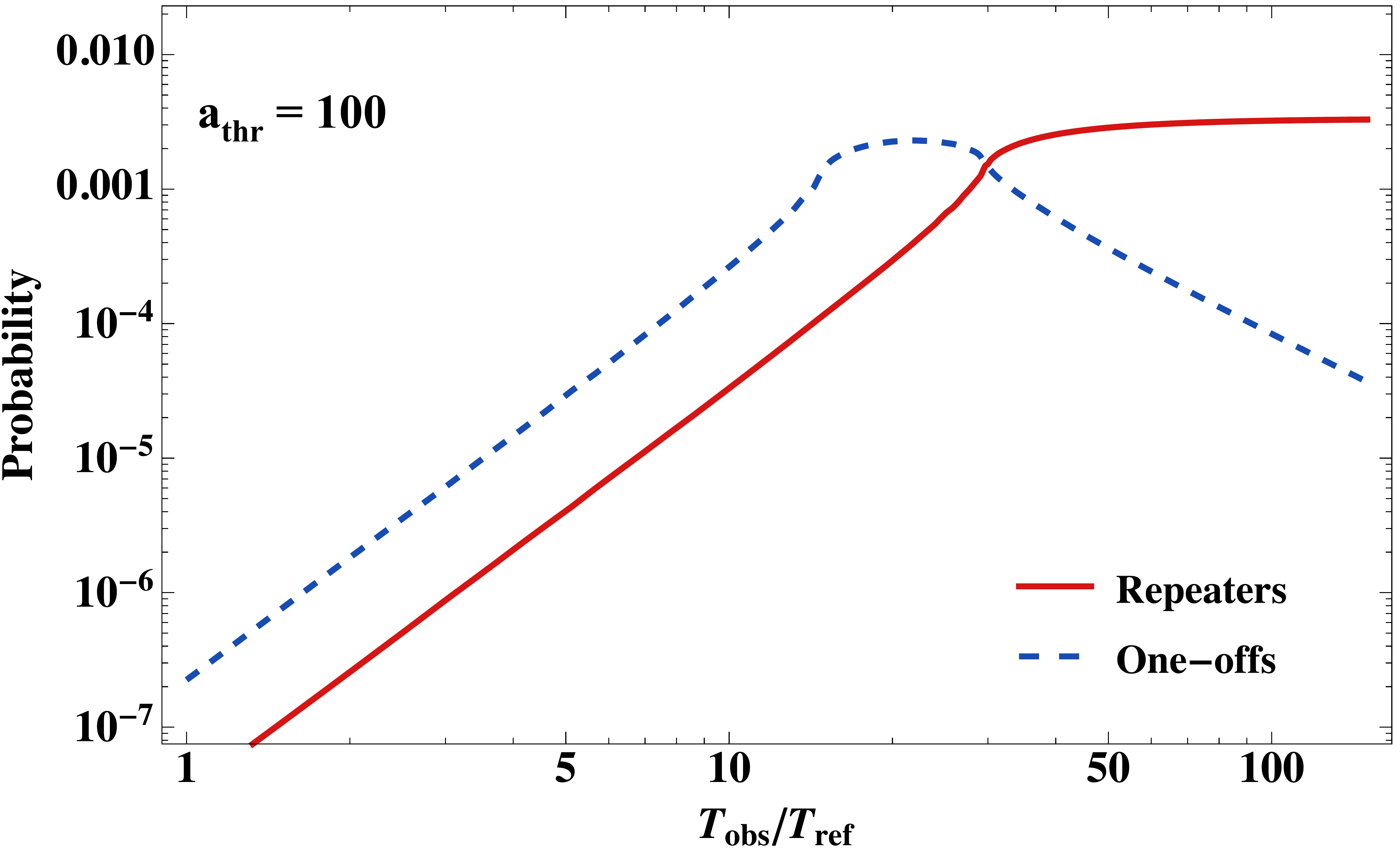}
\end{subfigure}
\caption{{\it Left Panels}:~Contours in the $(i, \xi)$-plane illustrating the parameter range for repeaters (red) and for one-off FRBs (blue), assuming $a_{\rm thr} =10$ (upper) or $a_{\rm thr}=100$ (lower), and $N_{\rm int} = 50$ (solid) or $N_{\rm int} = 150$ (dashed).~Values were chosen for illustration purposes.~For a given source, {\it i.e.} a fixed ${\cal R}$, increasing $N_{\rm int}$ is equivalent to assuming a longer observation time (see text).~In both plots the~dot-dashed horizontal green line separates the region of sources that can be amplified above the chosen $a_{\rm thr}$ (below the line), from those that cannot (above the line) and, thus, cannot be detected.~For illustration purposes the $y$-axis reports ($\xi-i$):~thus the locus $i = \xi$  lies along the $x$-axis while the locus $\xi = 180$\textdegree$ - i$ is the black diagonal on the right-hand side (only visible in the upper plot, due to a different scale on the $x$-axis in the lower plot).~For 
symmetry reasons, the $(i, \xi)$ combinations lying between these two loci are sufficient to describe the whole plane - {\it Right Panels:}~probability of obtaining a repeater (red solid) or a one-off FRB (blue dashed) vs.~observing time ($T_{\rm obs}$), for $a_{\rm thr} =10$ (upper) or 100 (lower).~For each $T_{\rm obs}$, we integrated the probability of individual ($i, \xi$)-combinations, with both angles randomly distributed, over the area in parameter space that allows for repeaters or one-off FRBs (left panels), respectively.~The reference time is T$_{\rm ref}  = N_{\rm int}/{\cal R}$, with $N_{\rm int} = 50$.}
\label{fig:prob-repvsoneoff-Tobs}
\end{figure*}
By construction, the largest amplification along the path~of the hotspot occurs at $\varphi =0$, then decreases as $\varphi$ grows towards $\pi$.~
At a given emission radius, the condition $a \ge a_{\rm thr}$ imposes $\theta \leq \theta_{\rm max} (r, a_{\rm thr})$, where $\theta_{\rm max}$ is obtained by solving eq.~(\ref{eq:aweaklens}).~In turn,~$\theta_{\rm max}$  selects 
a symmetric range of $\varphi$-values around $\varphi=0$, from $-\pi/N_{\rm int}$ to $+\pi/N_{\rm int}$.~We thus obtain via~eq.~(\ref{eq:theta}) 
\be
\label{eq:condition}
 \arccos \left(\cos i \cos \xi + \sin i \sin \xi \cos \displaystyle \frac{\pi}{N_{\rm int}}\right) \leq \theta_{\rm max} (r, a_{\rm thr}) \, ,
\ee
which defines the region, in the $(i, \xi)$ plane, where 
at least one of the $N_{\rm int}$ events overcomes the detection threshold. 

Repeaters will be those sources for which, given an intrinsic burst rate (${\cal R}$) and a particular orientation ($i, \xi$), two or more events will be observed in time~T$_{\rm obs}$.~Similarly, one-off sources will produce one observed event under the same conditions. Thus, we can formally define
\begin{eqnarray}
\label{eq:define-repeaters}
 \displaystyle \int_{a_{\rm thr}}^{a_{\rm max}} P(a; i, \xi) da & \equiv & {\cal I}_{a_{\rm thr}} \ge \displaystyle \frac{2}{{\cal R} f_b T_{\rm obs}}\, , 
 ~{\rm  repeaters}\nonumber \\
 1 \leq {\cal I}_{a_{\rm thr}} {\cal R}f_b  T_{\rm obs} &< & 2 \, ,~~~~~~~~~~~~~~~~~~~~~~~~~{\rm one-offs} \\
 {\cal I}_{a_{\rm thr}} {\cal R}f_b  T_{\rm obs} & < & 1\, ,~~~~~~~~~~~~~~~~~~~~~~~~~{\rm non-FRBs}\nonumber
\end{eqnarray}

\begin{figure*}
\centering
\begin{subfigure}{0.49\textwidth}
\centering
\includegraphics[height=0.605\linewidth]{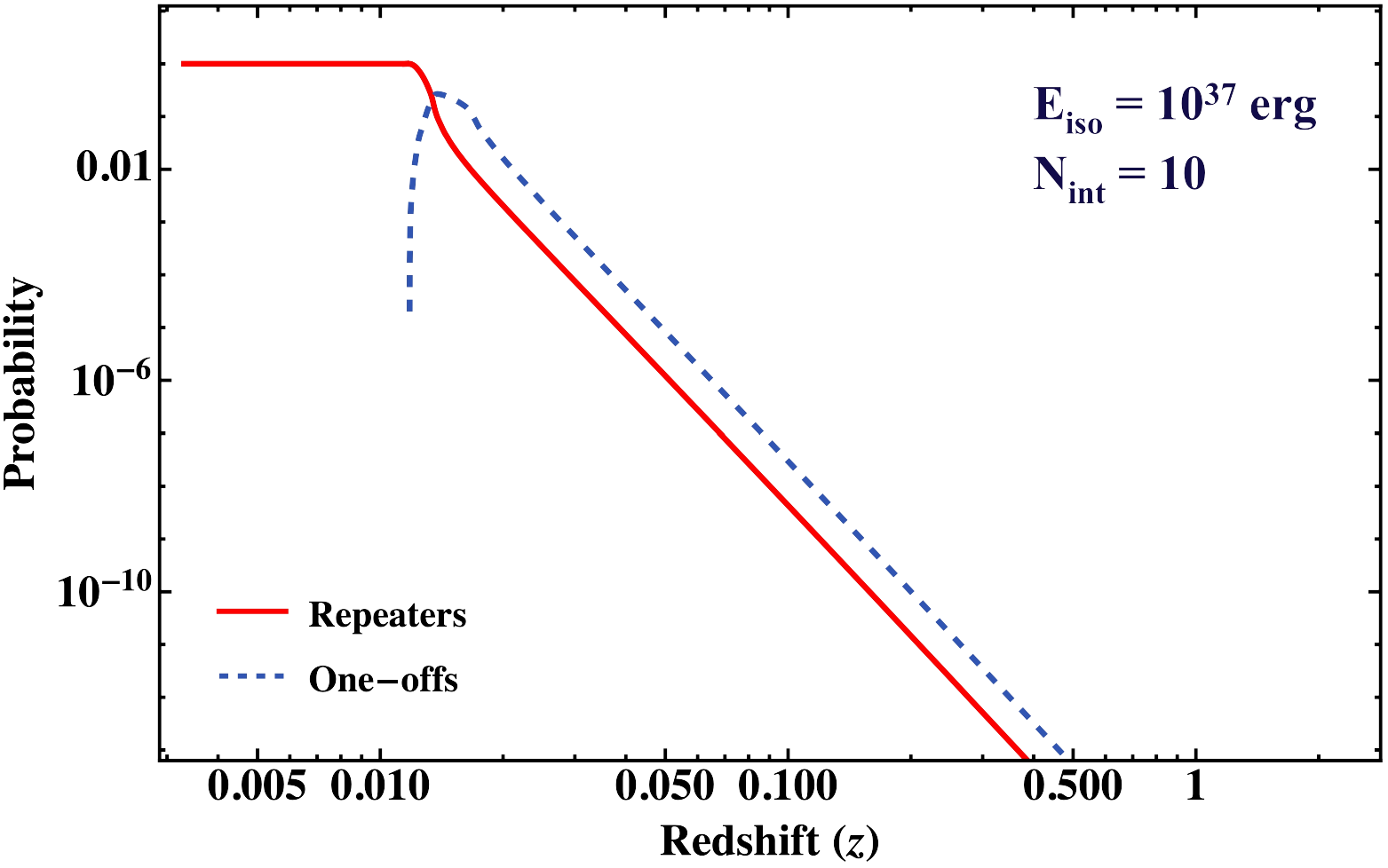}
\includegraphics[height=0.605\linewidth]{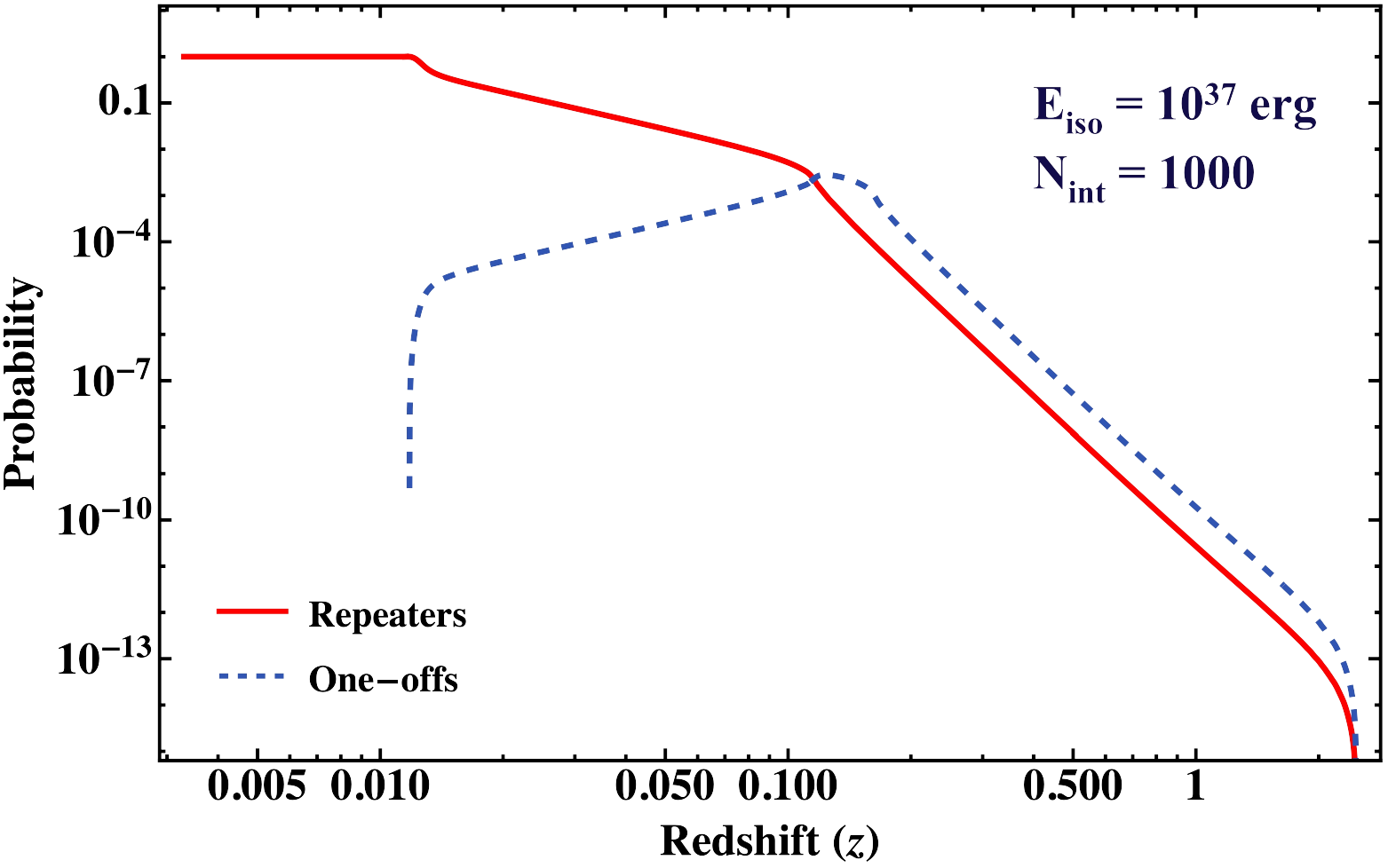}
\end{subfigure}
\begin{subfigure}{0.49\textwidth}
\centering
\includegraphics[height=0.605\linewidth]{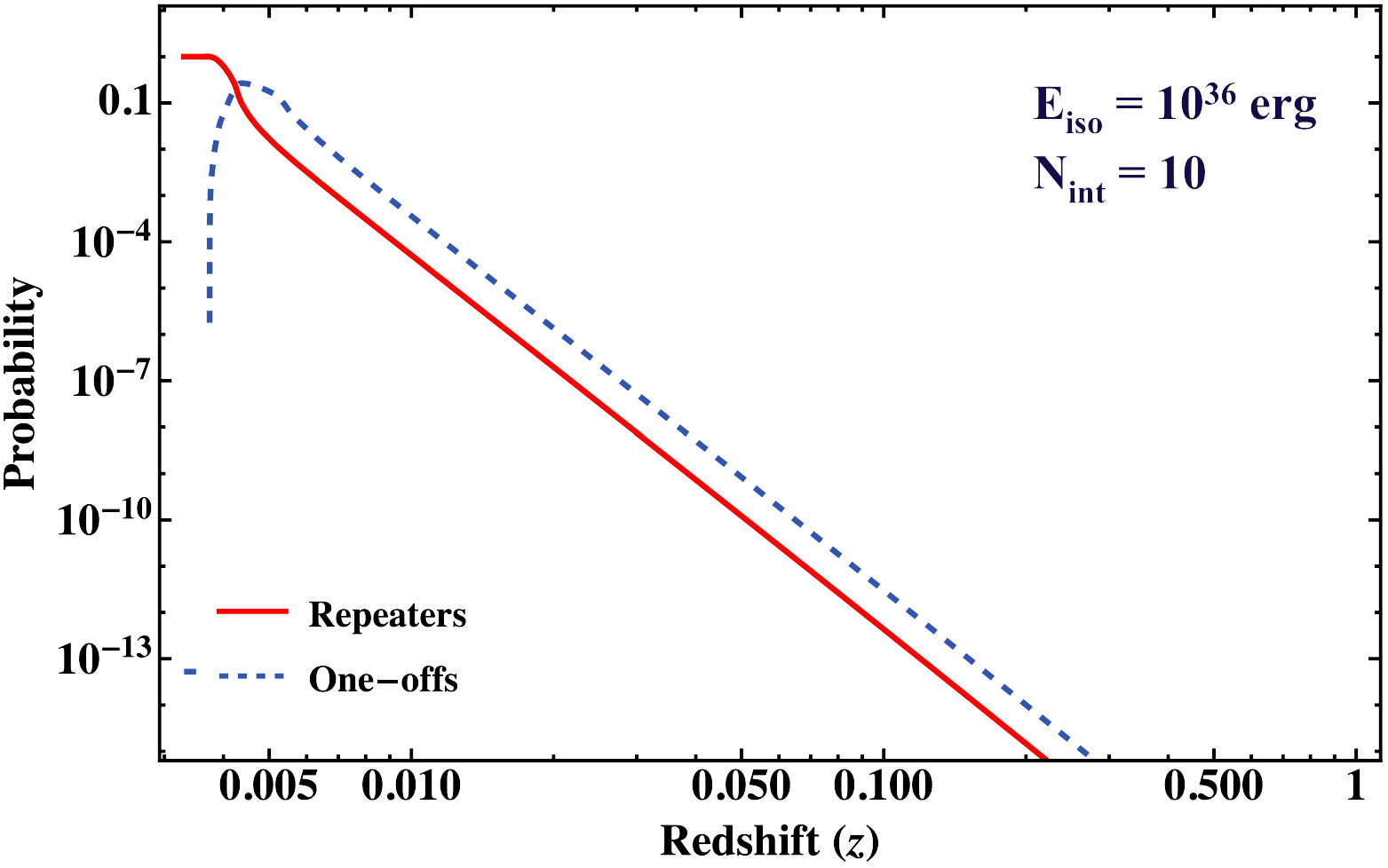}
\includegraphics[height=0.605\linewidth]{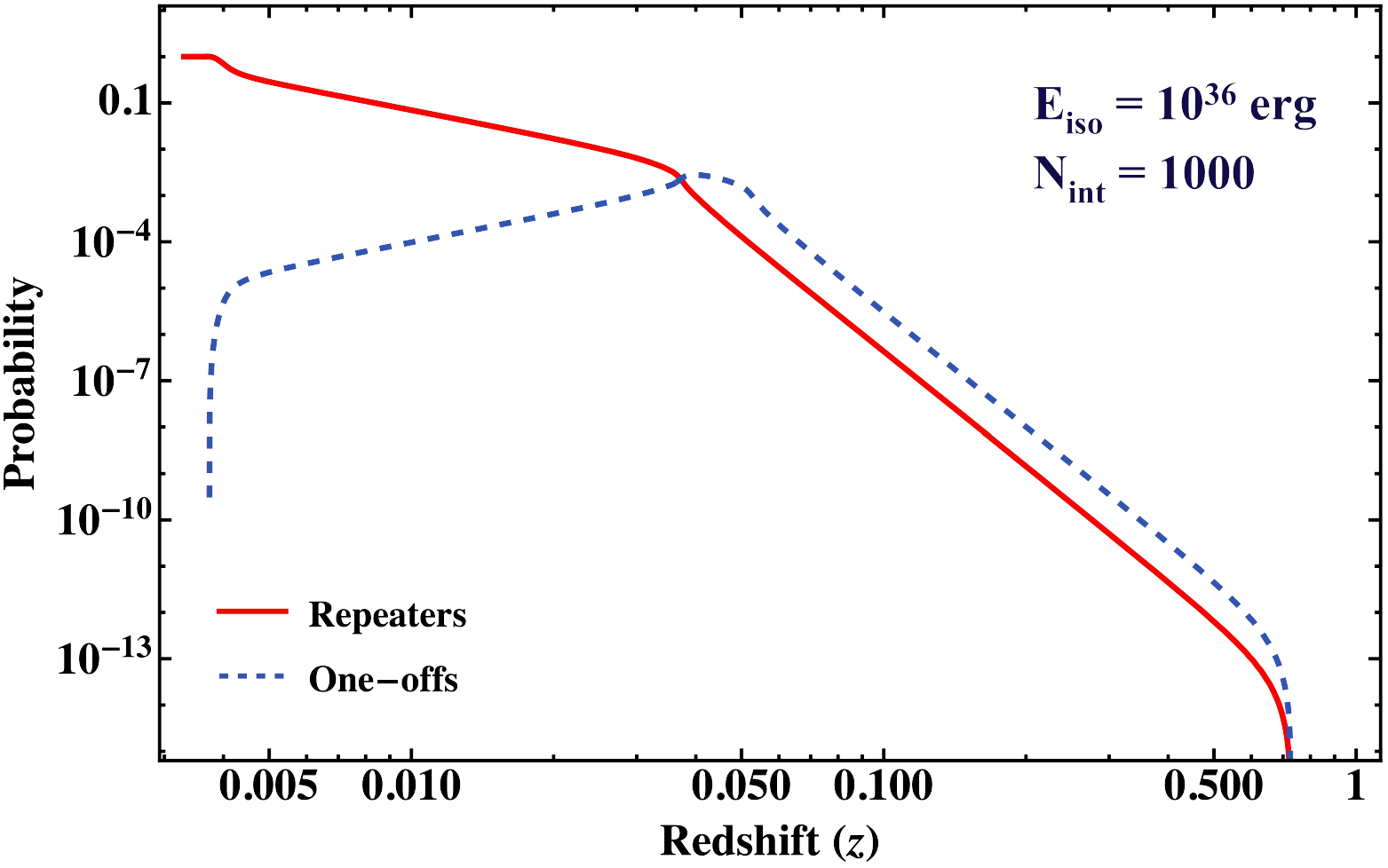}
\end{subfigure}
\caption{{\it Left Panels}: Probability of obtaining a repeater (solid, red) or a one-off FRB (dashed, blue) as a function of redshift, for standard candle 
hot-spots having $E_{\rm iso} = 10^{37}$ erg and total events $N_{\rm int} = {\cal R} f_b T_{\rm obs} = 10$ (upper) or $N_{\rm int} = 1000$ (lower).~The detection threshold was taken from CHIME at F$_{\rm thr}$ = 5 Jy ms \citep{li21}, corresponding to $a_{\rm thr} = 1$ at $z=0.145$ for the assumed standard candles - {\it Right Panels:} Same as left panels, but for 10 times weaker standard candles. The CHIME detection threshold corresponds, in this case, to $a_{\rm thr} = 10$ at $z=0.145$, and has $a_{\rm thr} =1$ at $z=0.047$.} 
\label{fig:Pdz-complete}
\end{figure*}
The integral ${\cal I}_{a_{\rm thr}}$ gives the probability that an event gets amplified above a given $a_{\rm thr}$, and~is uniquely determined by $a_{\rm thr}$ and the particular combination of $i$ and $\xi$.~For 
 a~given source, with a fixed set $(a_{\rm thr}; i, \xi)$, the parameter $N_{\rm int} = {\cal R} f_bT_{\rm obs}$ may thus be regarded as the minimum number of events required in time $T_{\rm obs}$, in order for a least one of them to be detected.

\subsection{Repeaters vs. non-repeaters vs. non-FRBs}
\label{sec:rep-vs-nonrep}
For a source producing $N_{\rm int}$ events in time T$_{\rm obs}$, and for~a given  threshold $a_{\rm thr}$, one can use the definitions of eq.~\ref{eq:define-repeaters} to find contours, in the $(i, \xi)$ plane, inside which at least one event or at least two events are expected, respectively.~The~latter contour defines the area in parameter space in which repeaters will occur, while one-off FRBs will fill the area~between the two contours.~Exterior to both contours is the area~of undetected sources.

The left panels in Fig.~\ref{fig:prob-repvsoneoff-Tobs} show such contours in the $i$~vs.~$\xi$ plane (repeaters in red, one-off FRBs in blue), 
for specific values of $a_{\rm thr} = 10$ (upper) or 100 (lower) and $N_{\rm int} = 50$ (solid) or 150 (dashed), chosen for illustration purposes.~Because $N_{\rm int} = {\cal R} f_b T_{\rm obs}$, changing the value of $N_{\rm int}$ for a given source (i.e. a fixed rate) is tantamount to changing the observation time.~Thus, the dashed contours in Fig.~\ref{fig:prob-repvsoneoff-Tobs} may be regarded~as the evolution of the solid ones, for the same source, as $T_{\rm obs}$ is increased by a factor 3.~For example, 
for a source with an intrinsic rate ${\cal R} = 1$ hr$^{-1}$, the values $N_{\rm int}=50(150)$ correspond to the observing times $T_{\rm obs} = 50(150)$ hrs.

We see that, as the observing time is increased, both the one-off and the repeater regions expand, but the latter grows faster, resulting in an increased probability for repeaters over one-off FRBs.~This expansion, though, hits a wall indicated by the green, dot-dashed horizontal line.~The combinations of $i$ and $\xi$ above this line cannot achieve amplifications $> a_{\rm thr}$:~thus, further increasing the observing time will not produce any new sources.~It will, however, eventually turn all one-off events into repeaters.~Note that most of the parameter space in Fig.~\ref{fig:prob-repvsoneoff-Tobs} corresponds to undetected sources (non-FRBs) which lie above the green-dashed line and, hence, will never become FRBs.

We also calculated the probability of observing repeaters~or one-off FRBs vs.~redshift for standard candle 
hot-spot
events.~To~this aim we defined a redshift-dependent~$a_{\rm thr}(z)$,
which depends on the standard candle's characteristic energy  ($E_{\rm iso}$) and on~the detector sensitivity, and then adopted the cosmological parameters measured by the 2018 Planck collaboration \citep[Table~2, last column in][]{planck20}.~For~the sake of our argument, we used the FAST fluence threshold, F$_{\rm thr} = 0.015$ Jy ms \citep{li21} and tried two different energies, E$_{{\rm iso}, 37} = 1$ or 0.1, which imply $a_{\rm thr}=10$ at $z\approx 0.61$ or at $z \approx 0.23$, respectively.~Moreover, we selected two widely different values of $N_{\rm int} = 10$ or 1000:~these can either represent sources observed for the same time but whose intrinsic rates differ by a factor 100, or sources with the same ${\cal R}$ but very different observing times.~With this set up we calculated, at each redshift, the contours shown in Fig.~\ref{fig:prob-repvsoneoff-Tobs}, and then integrated the probability of a generic ($i,\xi$)-combination, $1/4 \sin i \sin \xi$, over the parameter space corresponding to repeaters or one-off FRBs, respectively.
\begin{figure*}[ht]
\centering
\begin{subfigure}{0.49\textwidth}
\centering
\includegraphics[height=0.64\linewidth]{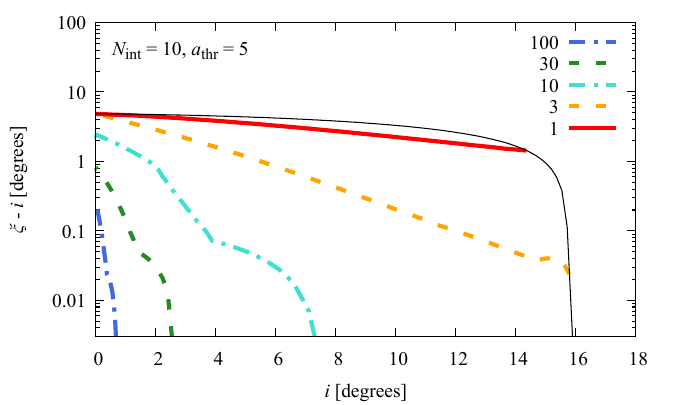}
    \end{subfigure}
\begin{subfigure}{0.49\textwidth}
\centering
\includegraphics[height=0.64\linewidth]{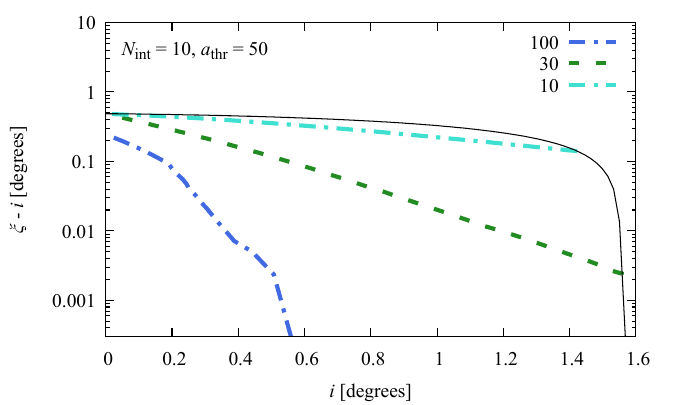}
    \end{subfigure}
    \begin{subfigure}{0.49\textwidth}
\centering
\includegraphics[height=0.64\linewidth]{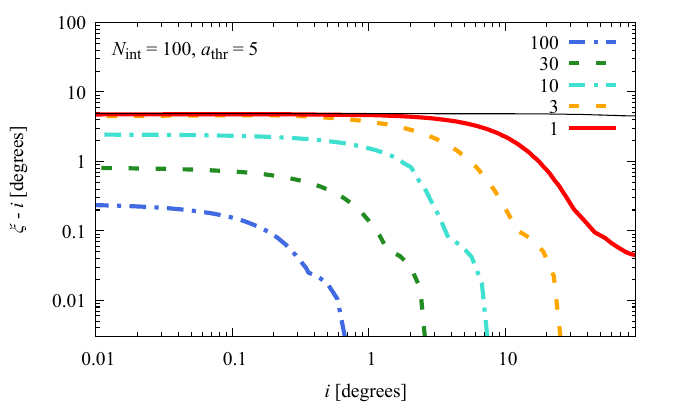}
    \end{subfigure}
    \begin{subfigure}{0.49\textwidth}
\centering
\includegraphics[height=0.64\linewidth]{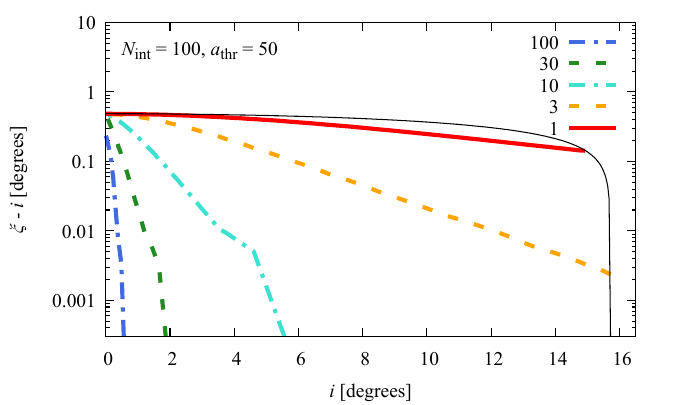}
    \end{subfigure}
\caption{Contours of the \textit{lensing gain} $ {\cal C} = \dot{E}_{\rm app}/\dot{E}_{\rm true}$, for standard candles at a fixed distance, as a function of the  angles~$i$~and $\xi$.~The four cases shown are for (a) amplification threshold $a_{\rm thr} = 5$ and $N_{\rm int} = 10$, representing a low-threshold and low-rate case, (b)~$a_{\rm thr} = 50$ and $N_{\rm int} = 10$, a high-threshold and low-rate case, (c) $a_{\rm thr} = 5$ and $N_{\rm int} = 100$, a low-threshold and high-rate case and (d) $a_{\rm thr} = 50$ and $N_{\rm int} = 100$, a high-threshold and high-rate case.~Configurations below the solid red curve have a lower power release than the apparent one because most of their flares are amplified and detected, {\it i.e.} the probability of exceeding $a_{\rm thr}$~is~high during most of the rotation.~Also shown are contours for an apparent-to-true power ratio of~3,~10,~30~or~100.~These~latter~finely-tuned configurations correspond to strong repeaters:~both their line of sight and emitting hot-spot lie close to the rotation axis, with very similar offsets.~Their probability of occurrence by random distributions of $i$ and $\xi$ is very low.~For reference, we calculated the probability of sources with ${\cal C} \geq 10$ and obtained, respectively $1.2 \times 10^{-6}$, $2\times 10^{-8}$, $1.2 \times 10^{-6}$, $8.7 \times 10^{-8}$ for cases (a)-(d).
}
\label{fig:energies}
\end{figure*}

Fig.~\ref{fig:Pdz-complete} depicts the resulting probability for sources of either type versus redshift, assuming a population of standard candles, 
for four combinations of $a_{\rm thr}$ and $N_{\rm int}$.~The probability of repeaters is maximal ($\leq 1$) at small $z$ and drops off steeply at $z \sim 0.015-0.1$ (depending on the adopted parameters).~Just beyond the same $z$-values, the probability of one-off events has a pronounced peak, then drops off with the same steep slope of repeaters.
~As a result, the one-off-to-repeater probability ratio stays constant ($\sim$ 8) at larger $z$, which is a specific expectation of the model\footnote{The exact value of their probability ratio, though, may be affected by further details not addressed in our standard candle approximation.}.~A lower energy of the events 
 shifts the probability profiles to lower $z$, maintaining their overall shape.~A low number of events ({\it i.e.} a shorter observing time or a lower burst rate) suppresses the one-off distribution at small redshift, and shifts both distributions to lower $z$. 
 
\subsection{Apparent vs.~true energy budget}
\label{sec:energybudget} 
Another crucial implication of gravitational lensing is that observed events should have a lower intrinsic energy.~In most configurations, significant amplifications will have very low probabilities, 
in turn implying a large number of non-amplified (undetected) events.~The combination of these two factors determines a substantial difference between the observed (or ``apparent'') energy budget of a source and its emitted one.~In extremely rare combinations of $i$ and $\xi$, though, significant amplifications may be achieved with a large probability 
and the source will appear as a very active repeater.~In such cases the apparent energy budget of a source will exceed, 
sometimes by a large factor, its true energy budget.
This is a specific and key expectation of our model.
In our simple assumption of standard candle hot-spot events, we calculated the ratio of apparent to true power release of~a given source as a function of $i$ and $\xi$, for two fixed amplification (detection) thresholds, $a_{\rm thr} = 5$ or 50,~and for two different values of $N_{\rm int}=10$ and $100$.~We label this ratio, ${\cal C} = \dot{E}_{\rm app}/\dot{E}_{\rm true}$, the {\it lensing gain}~and provide details of its calculation in Appendix \ref{app:standard} (Eq.~\ref{eq:ratio}).~Fig.~\ref{fig:energies} illustrates~our results, highlighting the region in parameter space where ${\cal C} >1$  (see caption for details).~For 
each of the four panels, we calculated the probability of configurations with a gain ${\cal C} \geq 10$, assuming a random distribution of both angles $i$ and $\xi$:~the two panels on the left have nearly coincident probabilities $\approx 10^{-6}$, the upper right panel has a probability $\approx 8.7 \times 10^{-8}$ and the lower right panel of $\approx 2\times 10^{-8}$.~In all cases, these are  
extremely unlikely configurations, 
implying a large population of sources to explain the 
existence of very active repeaters.~Note that in sect.~\ref{sec:implications} we estimated for FRB 20121102A and FRB 20201124A a probability $\sim 10^{-6}$ and a combination of angles which would correspond to ${\cal C} \geq 10$, e.g. in the left panels of Fig.~\ref{fig:energies}.~A detailed characterization of the active repeaters will be presented in an accompanying paper \citep{LaPlaca2024}.~Finally, 
Tab.~\ref{tab:simultab} shows a few examples for selected values of $a_{\rm thr}, i, \xi$.

\section{Gravitational lensing of a cosmic population of sources}
\label{ssec:popstudy}

\begin{figure*}
\centering
\begin{subfigure}{0.495\textwidth}
\centering
\includegraphics[width=\linewidth, height=0.65\linewidth]{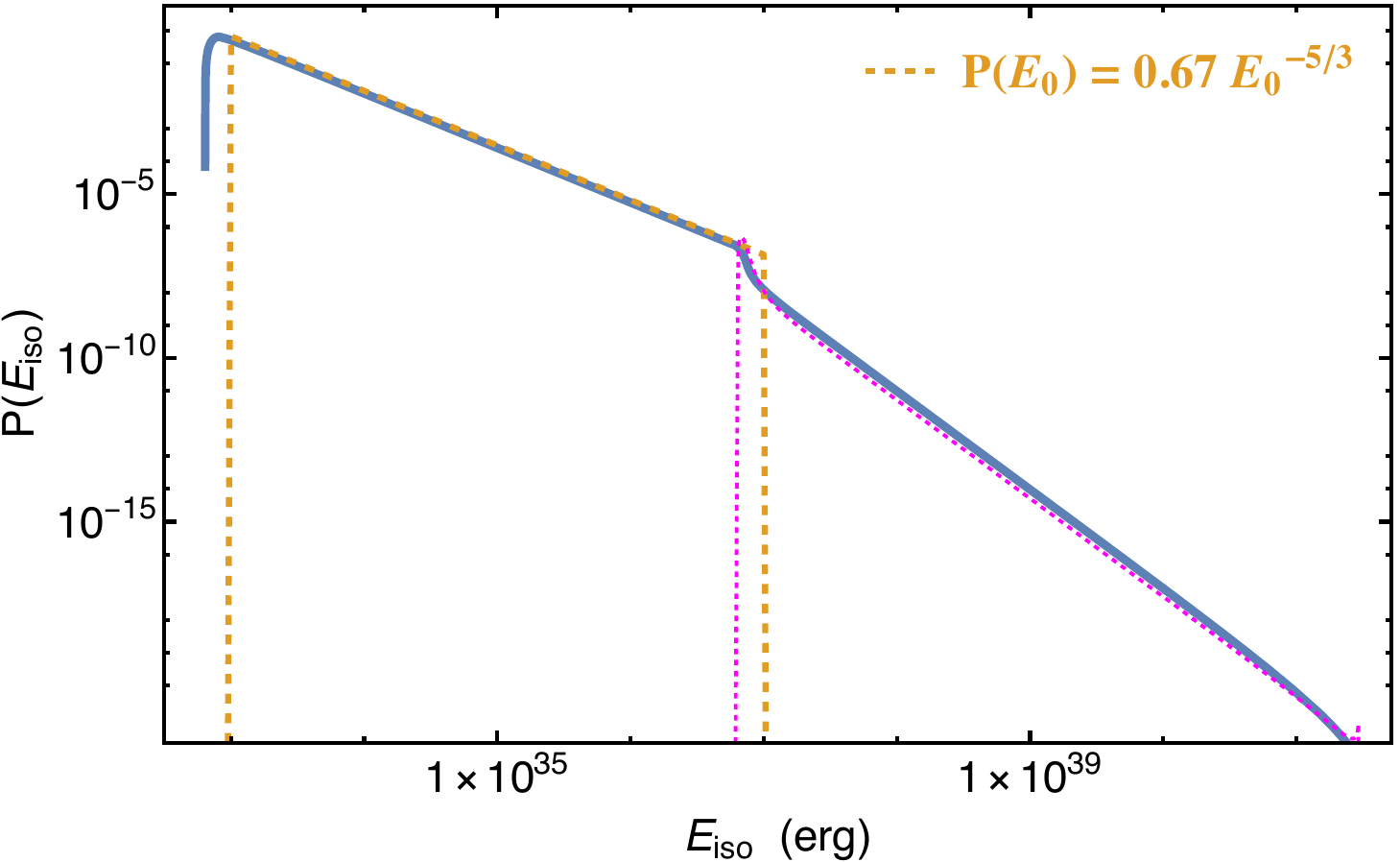}
\end{subfigure}
\begin{subfigure}{0.495\textwidth}
\centering
\includegraphics[width=\linewidth, height=0.65\linewidth]%
{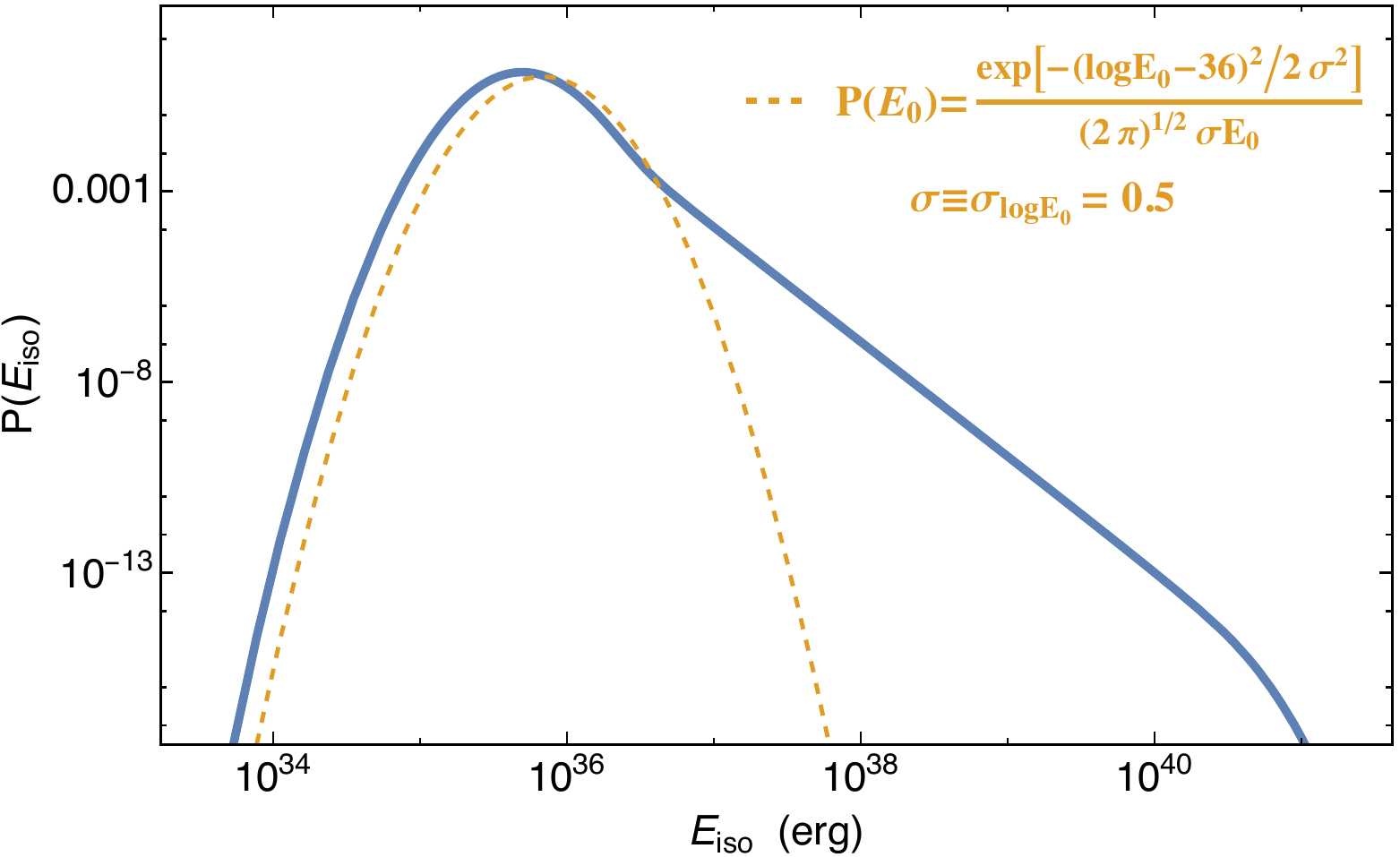}
\end{subfigure}
\centering
\begin{subfigure}{0.495\textwidth}
\centering
\includegraphics[width=\linewidth, height=0.44\linewidth]{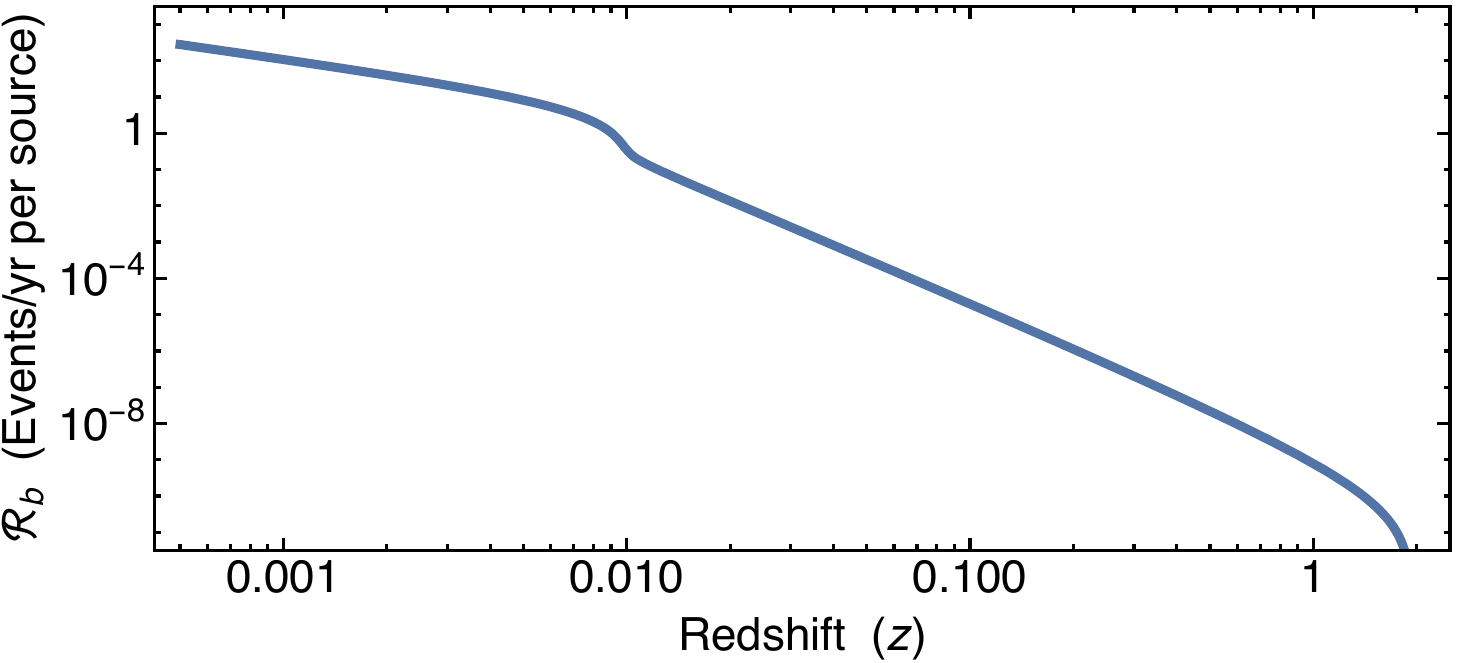}
\end{subfigure}
\centering
\begin{subfigure}{0.495\textwidth}
\centering
\includegraphics[width=\linewidth, height=0.44\linewidth]{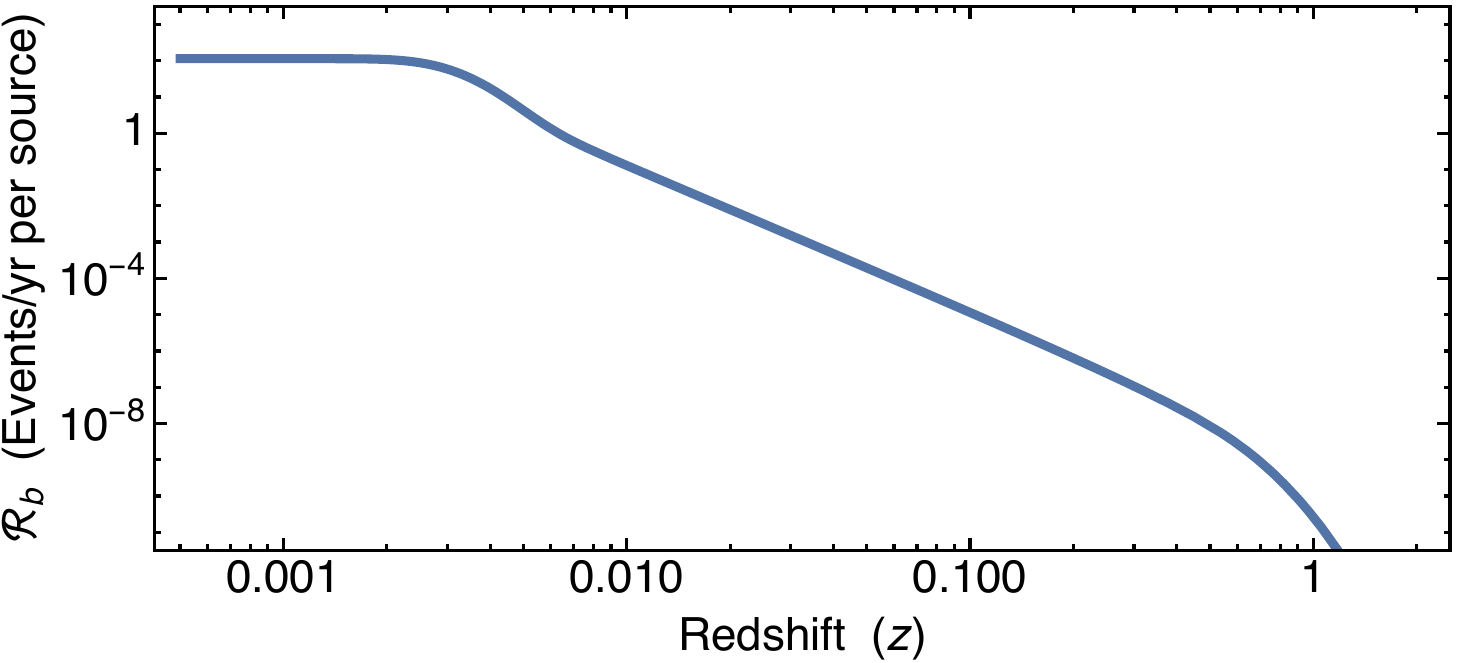}
\end{subfigure}
\caption{Predicted energy distribution of observed events, $\hat{P}(E)$ (blue, solid), for two different seed distributions (orange, short-dashed curves). {\it Upper left}: power-law seed, $P(E_0) \propto E_0^{-5/3}$, from $10^{33}$~erg to $10^{37}~{\rm erg}$, with superposed the function $\hat{P}(a)$ (magenta, dotted) producing a high-energy tail $\propto E^{-3}$ and a gradual transition in the range $0.64 \leq a <1 $, due to gravitational redshift; {\it Upper right panel}:~log-normal seed, with expectation value $E_{0, {\rm av}} = 10^{36}~{\rm erg}$ and $\sigma_{E_0} = 0.5$.~At high energies, lensing produces the power-law tail $\propto E^{-3}$.~At low energies, $\hat{P}(E)$ is offset relative to the seed, due to gravitational redshift; {\it Lower panels}:~The (average) rate of {\it observed} events per yr per source  ($R_b ={\cal N}_b/T_b$) vs.~redshift for each of the seed distributions. The two curves give very similar rates, with the power-law case declining earlier at low-$z$ but remaining somewhat larger at high redshifts.~Note that the event rate per source is $<10^{-4}$ yr$^{-1}$ (the estimated average rate per magnetar of eq.~\ref{eq:estimate-rate}) at $z \gtrsim 0.05$:~the all-sky rate~is thus dominated by local sources, and the observations of FRBs at high redshifts requires the existence of a large source population (e.g. at least $10^8$ sources at $z\approx 0.5$, in line with the estimated number assuming $T_b \approx 10^4$ yrs, in order to observe one event from that redshift in $T_{\rm obs}=1$ yr).}
\label{fig:pdiacap}
\end{figure*}

In Sec.~\ref{sec:repeaters}, \ref{sec:rep-vs-nonrep} and  \ref{sec:energybudget}, we only considered standard candle 
hotspots ~in order to highlight the role of lensing and of geometrical effects in shaping their possible appearence.~In such a population of identical 
hotspots, the observed energy distribution will directly track $\hat{P}(a)$, as discussed in sec.~\ref{sec:obsvsint}.~Instead, 
hotspots are likely to have an intrinsic (``seed'') energy distribution, $P(E_0)$, in which case the observed $\hat{P}(E)$ will result from the convolution of $\hat{P}(a)$~with that seed (eq.~\ref{eq:PdLbis}).~Here we 
considered two possible seeds (i) a power-law $P(E_0) \propto~E_0^{-5/3}$, inspired by the known energy distribution of magnetar X-ray bursts/flares, ranging from $E_{0, \rm min} = 10^{33}$~erg to $E_{0, \rm max} = 10^{37}~{\rm erg}$ and~(ii)~a log-normal, in analogy to the low-energy distribution observed in FRB 20121102A and 20201124A and to the typical pulses of radio pulsars, centered at $E_{0, {\rm c}} = 10^{36}~{\rm erg}$ and with $\sigma_{\log E_0}=0.5$.~As~in previous sections, we calculate the resulting $\hat{P}(E)$ for hotspot of linear size $\ell = 30$~cm at $R=10~r_g$.~Our results are illustrated in Fig.~\ref{fig:pdiacap}:~for both seed distributions, lensing extends significantly the high-energy tail, shaping it into the characteristic power-law with index $\lesssim 3$ (more details are provided in the caption).
\subsection{Impact of cosmology}
The energy distribution of the observed population of FRBs will be affected by additional factors, primarily the sensitivity of the detector 
and the non-uniform spatial distribution of sources.~The former introduces a $z$-dependent minimum energy, $E_{\rm thr}(z)$, in the observable $\hat{P}(E)$,
selecting an increasingly higher-energy tail of events at higher $z$.~As~a result, the average number of observed bursts per source at redshift $z$ is  
\be \label{eq:nbz}
N_b(z) = {\cal R} f_b T_b
\int_{E_{\rm thr(z)}}^{E_{\rm max}} \hat{P}(E) dE \, ,
\ee 
where we still assumed a unique ${\cal R}$ for all 
sources.~The maximum observed energy, $E_{\rm max}$, is $E_{0, {\rm max}}$ times~the maximum amplification ($a_{\rm max}$) and is independent of $z$.~The number of observed events per source per yr, ${\cal N}_b(z)/T_b$ for either seed distribution is depicted in the lower panels  of Fig.~\ref{fig:pdiacap}.

The spatial distribution of sources, on the other hand, determines their total number, hence the total number of detected events, at each redshift.~We assume that 
FRB~sources trace~the star formation history of the universe, e.g. they are associated to young NS\footnote{This is supported, for $z \gtrsim 0.05$, by recent studies \citep[]{zhang22} which also found hints of an additional population in the local universe.} with age $\le~T_b$. The NS birth~rate~per unit volume 
is related to the star formation rate (SFR) $\psi(z)$ through 
\be 
\dot{n}_{\rm NS}(z) = \psi(z) \nu_{\rm NS} \, ,
\ee 
where $\nu_{\rm NS} \approx 0.0068$ M$_\odot^{-1}$ is the number of NSs formed per unit stellar mass, taken to be independent on $z$, and the SFR is (\citealt{madau14})
\be \label{eq:SFR}
\psi(z) = 0.015\displaystyle\frac{(1+z)^{2.7}}{1 + \left[(1+z)/2.9\right]^{5.6}} ~ \displaystyle\frac{M_{\odot}}{\rm yr ~ Mpc^3} \, .
\ee
The rate of observable events up to redshift $z_{\rm max}$~results by multiplying $\dot{n}_{\rm NS}(z)/(1+z)$, the NS birth rate per unit volume\footnote{The $(1+z)$ term accounts for cosmological time dilation.}, by the number of observable events per source at that redshift, $N_b(z)$, and integrating over cosmological volume.  
\begin{table}
\begin{center}
\caption{Ratio of apparent to true power release (lensing gain) for three  $a_{\rm thr}$-values~and a few selected ($i, \xi$)-combinations.~The last columns show 
the range of amplifications spanned by each configuration.}
\begin{tabular}{|c|c|c|c|c|c|c|}
\hline
$a_{\rm thr}$ & $i$ & $\xi$ & ${\cal C}$ & $P(a>a_{\rm thr})$ & $a_{\rm max}$ & $a_{\rm min}$\\
~& (deg) & (deg) & ($\dot{E}_{\rm app}/\dot{E}_{\rm true}$) & (${\cal I}_{\rm athr}$) & & \\
\hline
\hline
2   & 3 & 3.01 & 20 &  1 & 2437 & 4.05\\
2   & 30 & 30.01 & 2.1 & 0.137 &2437  & 0.76 \\
2 & 3 & 4 & 7.4 & 1 & 24.4 & 3.48\\
2 & 30 & 31 & 0.85 & 0.133& 24.4 & 0.76\\
\hline
     10  & 3 & 3.01  & 16.1 & 0.263  & 2437 & 4.05\\
     10 & 30 & 30.01 & 1.7 & 0.027 & 2437 & 0.76 \\
      10 & 3 & 4 & 3.5 & 0.208 & 24.4 & 3.48\\
      10 & 30 & 31 & 0.4 & 0.024 & 24.4 & 0.76\\
      \hline
      100 & 3 & 3.01 & 10.0 & 0.026 & 2437 & 4.05\\
      100 & 30 & 30.01 & 1.05 & 0.0027 & 2437  & 0.76\\
        100 & 3 & 3.2 & 1.6 & 0.014 & 121.9 & 3.93\\
      100 & 30 & 30.2 & 0.18 & 0.0015 & 121.9 & 0.76\\
\hline
\end{tabular}
\label{tab:simultab}
\end{center}
\end{table}
The only two unknowns in this calculation
are the value of ${\cal R} T_b f_b$ in Eq. \ref{eq:nbz}, {\it i.e.}~the normalization of $N_b(z)$, and the calibration of $E_{\rm thr}(z)$, which is set by the detector sensitivity.~For a fluence-limited detector, we translate redshifts to energy thresholds through \citep[see, e.g.,][]{zhang18}
\be \label{eq:ethrz}
E_{\rm thr}(z) = {\cal{F}}_{\nu}^{\rm thr}~\nu_c~\frac{4 \pi D^2_L(z)}{1+z}  \, , 
\ee
where ${\cal{F}}_{\nu}^{\rm thr}$ is the detector fluence threshold, $\nu_c$ the central observing frequency, and $D_L(z)$ the luminosity distance. We considered here ${\cal{F}}_{\nu}^{\rm thr} = 5$~Jy~ms, which is the~CHIME 90\%-completeness fluence limit 
\citep[][]{nan11,li17frb}.

The value of ${\cal R} T_b f_b$ is set by requiring that the calculated all-sky rate of observable events equals 
the actual one.
\subsection[Results: cosmic populations with a universal rate]{Results: cosmic populations with a universal rate ${\cal R}$}
\label{sec:results}
With the above prescriptions, we evaluated the properties of the source population implied by our model under the simple assumption of identical sources, each emitting bursts~with the same seed energy distribution, either a power-law~or a log-normal, as described above.~Again we  assumed~that $\sim$~10\%~of all NS, i.e. magnetars, 
are FRB sources, and obtained that:

(i)~for the power-law seed distribution, the average number of events per source is ${\cal R} T_b f_b \sim 2 \times 10^7$ over the active lifetime $T_b$.~To estimate the overall energy budget of each source, we assumed~$\epsilon_r~\sim~10^{-5}$ for the unknown radio emission efficiency, 
comparable to the measured radio-to-X-ray ratio in the brightest FRB emitted by SGR 1935+2154.
~This assumes an X-ray efficiency of order unity in that event,~and implies an average energy budget $E_{\rm tot} \sim 10^{47}$~ergs per source~in FRB-related events, 
and ${\cal R} \sim 6$~day$^{-1} (10^4~{\rm yrs}/T_b)$.~The inferred value of ${\cal R}$ scales with the average energy per burst $\propto E_{\rm av}^{-3/2}$ ($E_{\rm av} \sim 5\times10^{34}$ ergs with our assumptions), while $E_{\rm tot} \propto E^{-1}_{\rm av}$.~Moreover, only $\sim 10^3$~of the $10^7$ events are~in~the amplified tail above $10^{37}$~ergs (1 every 10 yrs).~Among them, only a few events amplified above $a_{\rm thr}(z)$ - as set by the detector's sensitivity - will eventually be detected;

(ii)~for~the log-normal seed, the number of events per source is reduced to ${\cal R} T_b f_b \sim 10^6$, as their average energy is larger ($10^{36}$ ergs), 
the energy budget is roughly the same, $E_{\rm tot} \sim 10^{47}$~ergs, and the event rate is lower, ${\cal R} \sim 0.3$~day$^{-1}$.~Also~in this case, a minority of the events ($\approx 4500$) are in the amplified tail at $E_{\rm iso} > 4 \times 10^{36}$ ergs (see Fig.~\ref{fig:pdiacap}), or one every 2 yrs.

For both seed distributions, the lower panels in Fig.~\ref{fig:pdiacap}
clarify that the expected number of observed events per source is much lower than unity at $z >0.01$ (300 Mpc).~Thus, detecting a source beyond the local universe is intrinsically unlikely,~and only the 
very 
large number of NSs
at increasing redshifts
guarantees that some events will be observed out to~$z\lesssim 0.5$ (see caption).

The paucity of FRB sources in the local universe, where detection is in principle possible without lensing, places an additional constraint on their population, with the caveat that selection criteria for the CHIME catalog are likely to reject a fraction of nearby sources, i.e.~those with a DM $<$ 1.5~times the galactic value.~In Appendix \ref{app:C} we provide some illustrative examples on how this issue can be addressed, and the important conclusions that can be drawn.~More detailed 
results and constraints 
based 
on the proposed scenario will require a thorough population synthesis study, which is deferred to a future work.

\section{Summary and Conclusions}
We presented a new scenario for FRBs generated~in NS magnetospheres and amplified, via    
gravitational self-lensing in the strong field limit, by the~NS~itself.~We developed a semi-analytical approximation of the complex dependence of the amplification $a(\theta)$ for all values of $\theta$ in different lensing regimes, in order to obtain a numerically accurate and straightforward calculation of the amplification probabilities in different geometries (see
sec.~\ref{sec:Pdia}).~Events~occurring close to the caustic line can be amplified in intensity~by several orders of magnitude, thus alleviating the energy cost~of individual bursts without invoking highly relativistic motion.~Large 
amplifications
occur with low probability, implying~that cosmological FRBs represent the tip of the iceberg of a very wide population of undetected events.~In particular, if~detected FRBs are only those events amplified above a minimal~threshold ($a_{\rm thr}$), the relation ${\cal R}_b = f_b  P(a>a_{\rm thr}) {\cal R}$ holds between the rate of detections from a given source~(${\cal R}_b$) and its intrinsic event rate (${\cal R}$), where $P(a> a_{\rm thr})$ represents the probability of exceeding the minimal threshold,~and $f_b <1$ the emission beaming factor. 

~To 
determine $P(a)$, we explored~the possibility that~bursts occur either 
at random positions 
in an extended, spherically-symmetric region of the magnetosphere, or in a hot-spot corotating with the NS, thus sweeping an annular (ring-like) region.~For individual sources, the 
probability of large amplifications is a power-law $P(a) \propto a^{-n}$ with~$n=3$~in the spherically-symmetric case, and $n=2$~for~a ring-like emission region.~The energy distribution of~amplified events follows the same power-law, 
whose slope is consistent with that observed 
from individual well-studied sources (FRB 20121102A and FRB  20201124A); this lends support to the rotating hot-spot scenario.

For the population as a whole,~on the other hand, the randomly distributed viewing angles introduce spherical symmetry, thus favouring a power-law index $n= 3$ regardless of the emission geometry,~i.e.~even~if individual sources follow a flatter distribution.~Additional factors,~e.g.~a significant radial extension and/or angular structure~of the emitting region, or a physical constraint to the hot-spot latitude, can affect these slopes,~making them steeper (or flatter) at both the low- and high-amplification ends, in individual sources and in the population.~Through 
a few illustrative examples we showed that these results are robust against the additional details of the model that we introduced, as they are primarily a consequence of the behaviour of $a(\theta)$.

A key result of our scenario is that
the dichotomy of one-off FRBs and repeaters is found to emerge naturally in the rotating hot-spot scenario, 
provided that $a_{\rm thr} > 1$, as~a result of the extreme sensitivity of both the amplification~probability and range~to~the emission geometry.~For random distributions of the source orientation and hot-spot colatitude, and with the simple assumptions of  
standard candle hot-spots 
and identically-repeating sources, we find 
that (i) for observing times $T_{\rm obs} < 100 /{\cal R}$, the 
probability of a source appearing as a
repeating FRB is lower than for one-offs, and both probabilities are low within the underlying population, 
(ii) very active~repeaters correspond to the rarest cases in which most of the emitted~bursts are amplified and, thus, detected.~They 
require~both the line-of-sight and the hot-spot to lie at nearly identical and small angles ($\lesssim$ a few degrees) from the rotation axis:~thus, their probabilities are~$<10^{-5}$.~Since their properties make them easier to detect than more typical sources, very active repeaters will appear in a much larger proportion in FRB observations:~a quantitative assessment of this selection effect is, however, beyond the scope
of this paper and will be addressed in future work;
(iii)~as~$T_{\rm obs}$~is increased, the relative probability of repeaters steadily increases at the expense of one-off sources, which become increasingly more unlikely.

For identically-repeating sources emitting identical events with energy $E_0$, we evaluated the probability~of repeaters and one-off FRBs vs.~redshift, for different values of~$E_0$~and of $N_{\rm int} = f_b {\cal R} T_{\rm obs}$.~Adopting the CHIME detection threshold (${\cal F}_\nu^{\rm thr} \approx$ 5 Jy ms), we find that sources at~small distance, 
$D^2_L(z)/(1+z) <  {\rm a~few}\times E_0/(4\pi \nu_c {\cal F}_\nu^{\rm thr})\sim$10-100~Mpc for $E_0 = 10^{36-37}$ erg and $\nu_c = 1$~GHz,~do require little or no amplification for detection, hence they are likely to appear as repeaters if they emit at least a few bursts during $T_{\rm obs}$.~The repeater probability drops quickly beyond such distances, at first leading to a rise in the probability of one-off FRBs.~The latter, though, peaks in a narrow distance range beyond a few times $E_0/(4 \pi \nu_c {\cal F}_\nu^{\rm thr})$, and then shows a sharp drop at $z > 0.05-0.1$ (depending on the intrinsic energy $E_0$, and~on $N_{\rm int}$).~Beyond such redshifts, the one-off probability drops with the same slope of repeaters, remaining higher and at a constant ratio~$\sim7$ in our simple calculation.~The resulting probability of observing repeaters or one-off FRBs will be determined by a trade-off between its dependence on $T_{\rm obs}$ and the dependence on redshift.~Moreover, observational biases may further affect the observed population (e.g. \citealt{chime21, McGLor24}). 

The apparent energy budget of~individual sources in our scenario is typically smaller~than their true one.~However, in the ring-like geometry, we demonstrate that there is always a special subset of ``ideally-oriented'' systems in which we happen to observe most of the emitted flares with a significant amplification:~hence, their apparent energy budget is actually larger than their true one.~This is the subset to which the most~active repeaters, e.g. FRB 20121102 and FRB 20201124A, belong.~The area in the $(i, \xi)$-plane available for such active repeaters is determined by their intrinsic burst rate ${\cal R}$ and distance, and by our detection threshold, where a higher (lower) rate, smaller (larger) distance and lower (higher) threshold all favour a wider (smaller) 
area.~For the parameter range explored here, inspired by 
the characteristics of the above-mentioned repeaters, the probability of such configurations never exceeds $10^{-5}$~and can be much lower, particularly at larger redshifts.~An immediate implication is that, even for the huge source population of cosmological NSs, such active repeaters will appear in~very small numbers, and only out to a limited redshift. 

Finally, as a first step towards understanding the cosmological population, we studied the case of identical sources with the same rate ${\cal R}$, and in which events follow a (universal) seed energy distribution, e.g.~a power-law or a log-normal.~We assumed that young ($\le 10^4$ yrs of age) magnetars represent the cosmic population of FRB sources, which then tracks~the~cosmic SFR, and found they account well for 
the observed all-sky rate of FRBs requiring (i) a characteristic energy budget~of $10^{47}$~erg~per source, for a radio efficiency $\sim 10^{-5}$, perfectly~in line with the magnetic energy reservoir of magnetars and (ii) an intrinsic burst rate per source which is $100-10^3$ times~lower than observed in the most active repeaters. We note 
this is yet another peculiarity of the latter sources clearly indicating that, in order to match the properties of the FRB population, a realistic model will have to allow for a range of ${\cal R}$-values, reflecting a range of birth parameters and possibly their secular evolution.
We conclude that the neutron star self-lensing scenario introduced here provides a novel and promising avenue for interpreting many of the most relevant and yet elusive features of FRBs.

\vspace{-0.225cm}\section*{}
{\small This work was funded by the European Union’s~Horizon2020 research and innovation programme under the Marie Skłodowska-Curie (grant agreement No.754496).~SD also acknowledges support from the GSC at the Goethe University of Frankfurt, where this work was completed.~RLP acknowledges support from INAF's research grant {\it Uncovering the optical beat of the fastest magnetised neutron stars (FANS)} and from the Italian Ministry of University and Research (MUR), PRIN 2020 (prot. 2020BRP57Z) \textit{Gravitational and Electromagnetic-wave Sources in the Universe with current and next-generation detectors (GEMS)}.~AP acknowledges the contribution of the NextGenerationEU funds within the National Recovery and Resilience Plan (PNRR), Mission 4 - Education and Research, Component 2 - From Research to Business (M4C2), Investment Line 3.1 - {\it Strengthening and creation of Research Infrastructures, Project IR0000026 – Next Generation Croce del Nord}.~The visualization of functions in this manuscript was carried out with \texttt{Wolfram Inc. Mathematica}, \texttt{gnuplot}, and \texttt{GeoGebra}.}

\bibliographystyle{aasjournal}
\bibliography{bgraf}

\appendix

\section{Parametrization of lensing for a finite-size source}
\label{app:A}

In sec.~\ref{sec:Pdia} we presented an approximation (eq.~\ref{eq:aweaklens}) to the amplification of a point-like source which holds in all lensing regimes (extreme, strong and weak), here transcribed for ease of reference:
\be
\label{eq:wsapprox}
a(r,\theta) = 2 ~\displaystyle \frac{f(r)}{r \theta} ~ \left[1 + \displaystyle \left(\frac{\theta}{\theta_{\rm we}}\right)^{S}\right]^{1/S}\, .
\ee
Like eq.~(\ref{eq:apointlike}), which was presented in \citet{Bakala2023}, eq.~(\ref{eq:wsapprox}) is expressed as a function of the radial distance from the lens, $r$, and of the angular separation from the caustic, $\theta$, instead of the more common, in WDL studies, angular coordinate on the observer's sky.~Both eq.~(\ref{eq:apointlike}) and~(\ref{eq:wsapprox}) account for the first two images of the source, \textit{i.e.} the first direct image (FDI) and first indirect image (FII), since higher order images increasingly become negligible. 

\begin{figure}[b]
\centering
\includegraphics[height=0.28 \textwidth]{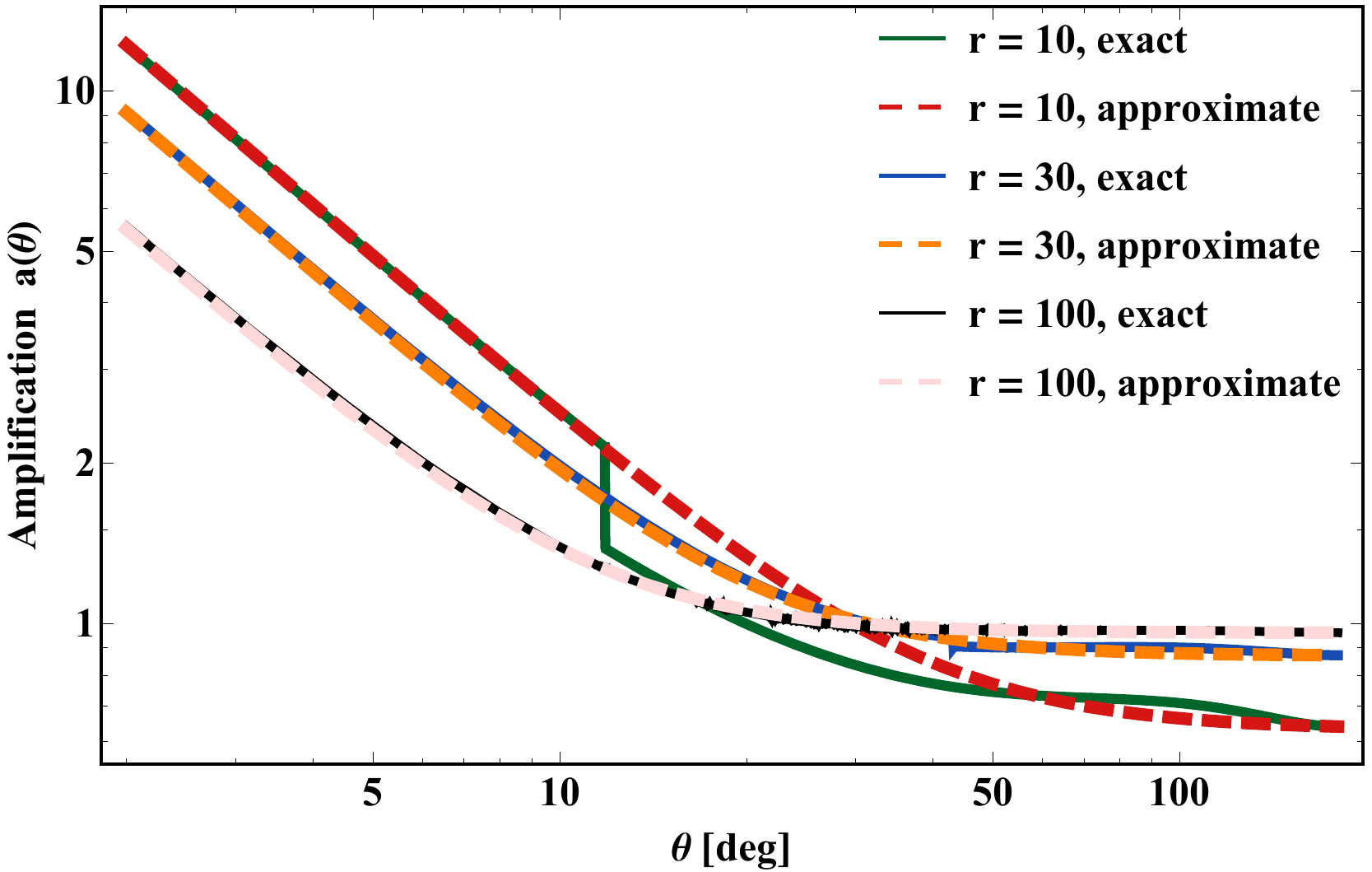} \hspace{0.5cm}
\includegraphics[height=0.28 \textwidth]{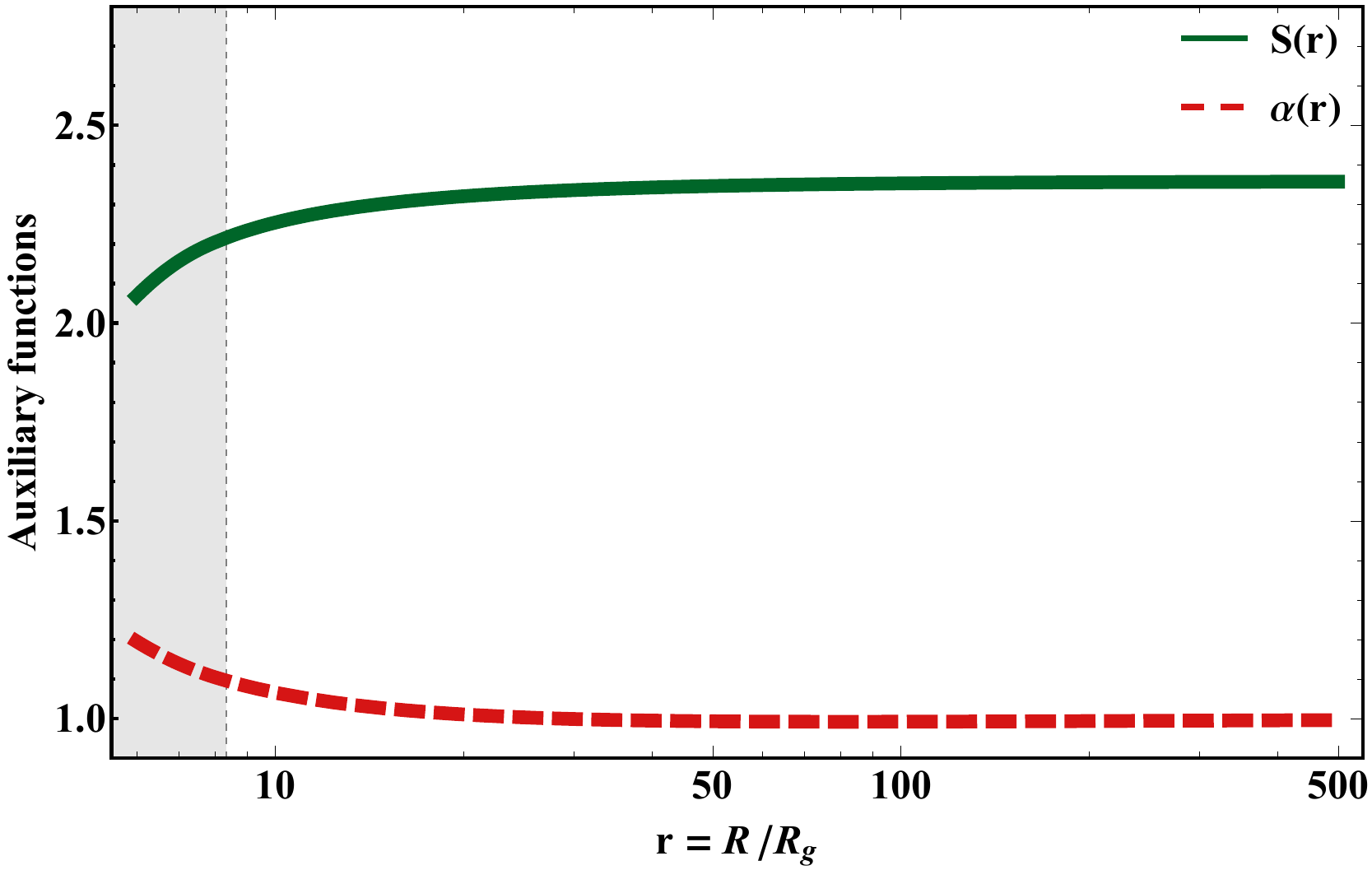}
\caption{{\it Left:}~The approximate expression of eq.~\ref{eq:wsapprox} for $a(\theta)$ (dashed curves) vs. the exact solution, for three different emission radii.~The step-like decrease of the green solid curve at intermediate $\theta$-values is due to the occultation of the FII by the NS surface (see main text)~{\it Right}:~Behaviour of the functions $\alpha(r)$ and $S(r)$ appearing in our approximations (eq. \ref{eq:wsapprox} and \ref{eq:complete-a}).~The gray shaded area represents sources for which even the FDI can be occulted by the NS surface.}
\label{fig:radialparam}
\end{figure}

At the close distances we are considering, $r \leq 100$, the conditions for the WDL are never fully satisfied, and the shape parameters $\theta_{\rm we}$ and $S$, both functions of $r$, are required to properly approximate the large-$\theta$ range, especially for NSs, as their surface can occult some of the  emission around them.~In particular, this happens to the FII of sources at small $r$ and close to $\theta= \pi$, i.e. exactly between the lens and the observer, which would in principle    
provide the main contribution to their amplification 
\citep[see, e.g., Fig.~2 in][]{Bakala2023}.~However, as $\theta$ grows, 
the periastron of the photon trajectory, i.e. the distance of closest approach to the center of the lens, decreases, eventually becoming smaller than the NS radius for these FIIs.~Therefore, their contribution 
must be discarded in all such cases.~Fig. \ref{fig:radialparam} 
shows the total amplification from the first two images as a function of $\theta$ for three different emission radii (solid lines, left panel), in the case of a NS with a $6~r_g$ radius: the sudden amplification drop at intermediate $\theta$-values is due to the occultation of the FII by the NS surface.~This effect is apparent for $r = 10$ (green solid curve), while it is already negligible at $ r \gtrsim 20~r_g$, as it takes place at larger values of $\theta$ and $r$, where the contribution of the FII quickly drops. 

In light of these considerations, we construct our approximated formula for $a(\theta)$ to be valid at all angular separations from the caustic for $r \gtrsim 10$ as a smoothly broken power-law: its derivative for $\theta \rightarrow \pi$ must go to zero to mimic the behaviour of the FDI, and its small-$\theta$ limit must follow the total amplification in the extreme lensing regime (eq. \ref{eq:apointlike}). This results in eq.~(\ref{eq:wsapprox}), in which the transition between the extreme and weak regime in amplification takes place at the angle $\theta_{\rm we}$, which at close radii will differ from the weak-strong transition angle 
$\psi_{\rm ws}(r) = 2/\sqrt{r}$, as calculated in 
the WDL\footnote{That is, where the angular coordinate of the source on the observer's sky corresponds to the apparent angular radius of the first Einstein ring.}: for ease of comparison with the WDL, we set $\theta_{\rm we}(r) \equiv \alpha(r)\,\psi_{\rm ws}(r)$.
The values of $\alpha(r)$ and $S(r)$ can be then found at any radius by imposing the following conditions:~(i) $a(r,\pi)~=~g^4 = \left(1-2/r\right)^2$ and (ii) $a(r, \theta_{\rm we})= a_{\rm WDL}\left(\psi_{\rm ws}(r)\right)=3 \sqrt{5}/5 \left(1-2/r\right)^2$, where $a_{\rm WDL}$ is the amplification in the WDL approximation \citep[see, e.g., sec. 3.1 in][]{Bakala2023}. We show the behaviour of the two parameters $S$ and $\alpha$ in the right panel of Fig.~\ref{fig:radialparam}: as expected, at larger radii the value of $\alpha$ tends to 1, indicating that the approximation tends to the WDL, while it increases at smaller radii, highlighting the deviation from the WDL in a stronger gravitational field.~For reference, we also shaded the region of the graph at emission radii lower than $r\simeq8.36$: below it, even the FDI trajectories can have a periastron below the NS surface \citep[for a star with radius $6~r_g$; see, e.g.,][]{LaPlaca2020}. 
Since the occultation of the FDI in the shaded region would cut out the high amplification regions behind the lens, sources at those radii are uninteresting for the purposes of our FRB model.

The resulting approximate amplification curves for three different radii (dashed lines) are compared to the exact solutions (solid lines) in the left panel of Fig. \ref{fig:radialparam}: the step-like decrease for the case of $r = 10$ mentioned above, is the only feature not modelled by our approximation, and does not impact on any of the conclusions in this work, as it mildly affects a region of relatively low amplifications. With that exception, the agreement of eq.~\ref{eq:wsapprox} with the exact solutions is excellent at all angular separations. 
\vspace{0.2cm}

In sec.~\ref{sec:finitesize} we introduced the saturation of amplifications, for sources with finite size $\ell$, at angular distances $\theta < \theta_s= \ell/r$ (eq.~\ref{eq:afinitesize}).~Combining that equation with (\ref{eq:wsapprox}), the amplification of finite-size sources in any lensing regime is well approximated~by 
\be
\label{eq:complete-a}
a_\ell(r,\theta) = 2 ~\displaystyle \frac{f(r)}{r \theta_s~\left[1+\left(\displaystyle \frac{\theta}{\theta_s}\right)^p\right]^{1/p}} \left[1 + \displaystyle \left(\frac{\theta}{\theta_{\rm we}}\right)^S\right]^{1/S}
\, ,
\ee
where we set the `smoothness parameter' $p=3$ to match the behaviour of the exact solutions at values of $\theta$ comparable to the angular size of the source \citep[see sec. 3.3 of][]{Bakala2023}. 

\section{Observed vs. intrinsic properties}
\label{app:pdevpda}
Here we elaborate further on the concepts introduced in Sec.~\ref{sec:obsvsint}, where we illustrated the crucial implication of gravitational lensing, {\it i.e.} the substantial difference between a source's apparent and real properties. 

\subsection{Luminosity function and source energetics for standard candle 
hot-spot 
events}
\label{app:standard}
In Sec.~\ref{sec:obsvsint} we showed that the 
probability distribution $P(S)$ translates to an observed rate  distribution, $F(S)$,
\be
\label{eq:defineFdLapp}
F(S) = {\cal R}_b P(S) = \displaystyle \frac{{\cal R}_b}{\hat{S}_0}~ \displaystyle \frac{P\left(S / \hat{S}_0\right)}{\displaystyle \int_{a_{\rm thr}}^{a_{\rm max}} P(a) da }\, ,
\ee
where ${\cal R}_b$ 
is the rate of {\it observed} events above the detection threshold and $P(a)$ is normalized such that $\int_{a_{\rm min}}^{a_{\rm max}} P(a) da =1$.~Here $a_{\rm max}$ is the highest possible amplification and $a_{\rm min}$ the minimum, typically $\lesssim 1$ (although special configurations may occur in which all events get amplified, \textit{i.e.}\ $a_{\rm min} >1$; see, e.g., Sec.~\ref{sec:rings}). 

The threshold amplification for detection, $a_{\rm thr}$, depends~on the intrinsic energy of the events, on the source distance,~and on the detector's sensitivity.~Hence, its value is not known a priori.~When all events from a source are observable,~\textit{i.e.} $a_{\rm thr} \leq a_{\rm min}$, then the observed event rate ${\cal R}_b$ will equal the intrinsic rate, ${\cal R}$.~More generically, though, only a fraction of bursts are expected to be amplified above threshold, hence $a_{\rm thr}~>~a_{\rm min}$ and ${\cal R}_b < {\cal R}$.~Accordingly, 
we may define an apparent source luminosity as
\be
\dot{E}_{\rm app} = \int_{{\rm S}_{\rm thr}}^{{\rm S}_{\rm max}} (f_b S) \left(\frac{F(S)}{f_b}\right) dS = {\cal R}_b \displaystyle \frac{\displaystyle \int_{{\rm S}_{\rm thr}}^{{\rm S}_{\rm max}} S \displaystyle \frac{P\left(S/\hat{S}_0\right)}{\hat{S}_0} dS}{\displaystyle \int_{a_{\rm thr}}^{a_{\rm max}} P(a) da } = {\cal R}_b \hat{S}_0 \displaystyle \frac{\displaystyle \int_{a_{\rm thr}}^{a_{\rm max}} a P(a) da}{\displaystyle \int_{a_{\rm thr}}^{a_{\rm max}} P(a) da } \, , 
\ee
where, by definition, $S= a \hat{S}_0$.~In the integrand, we have explicitly accounted for a possible beaming of the emission, multiplying the observed fluence by $f_b <1$, and dividing by $f_b$ the observed rate of bursts at fluence $S$.~This shows explicitly that beaming does not affect the estimated power of a source.~Along a similar line, 
the true source luminosity can be expressed as
\be
\label{eq:Etrue}
\dot{E}_{\rm true} =  \left(f_b \hat{S}_0\right) {\cal R} =  \frac{\displaystyle {\cal R}_b \hat{S}_0}  { \displaystyle \int_{a_{\rm thr}}^{a_{\rm max}} P(a) da} \, ,
\ee
where we adjusted the true isotropic fluence, $\hat{S}_0$, for~beaming, and converted ${\cal R}$ into the observed burst rate ${\cal R}_b$ (including beaming), according to the general relation
\be
\label{eq:rateb-vs-rate}
{\cal R}_b = f_b {\cal R}  \int_{a_{\rm thr}}^{a_{\rm max}} P(a) da \, ,
\ee
valid as long as the direction of beaming is random at each event, and the beaming angle is wider than the narrow range of directions over which lensed rays can reach the observer ($\delta \gamma \approx 10^{-4}$~deg).

Note that, contrary to the case of the source power, beaming does affect the estimated event rate of the source, ${\cal R}$.~Therefore, 
both lensing {\it and} beaming affect the estimated event rate of a source, but only lensing has an impact on its energy budget.~The latter can be easily quantified in this simple example, by writing the 
ratio between the apparent and actual power release
\be
\label{eq:ratio}
{\cal C} = \displaystyle \frac{\dot{E}_{\rm app}}{\dot{E}_{\rm true}} = \displaystyle \int_{a_{\rm thr}}^{a_{\rm max}} a P(a) da \, .
\ee
For standard candle 
hot-spots, ${\cal C}$ is simply defined by the range of observable amplifications and by the probability distribution $P(a)$.

\subsection{Events with an intrinsic luminosity function}
\label{app:lumfunc}
Here we generalize the calculation of the apparent and true source luminosity (power release) to the case in which the events have an intrinsic luminosity function, $P_0(S_0)$, at the source.~The apparent power is obtained by multiplying a given fluence by the rate of observed events at that fluence, and then integrating over all fluences, from the detection threshold to the maximum.~As for the true source power, we must first convert the intrinsic probability $P_0(S_0)$ into an intrinsic luminosity function, $F_0(S_0) = B P_0(S_0)$, by means of an expression similar to Eq.~\ref{eq:normalise}, only with ${\cal R}$ instead of ${\cal R}_b$.~We then proceed to integrate $F_0(S_0)$ between its minimum and maximum intrinsic values.~Hence
\be
\label{eq:energies}
\begin{cases}
\dot{E}_{\rm app} =\displaystyle  \int_{S_{\rm thr}}^{S_{\rm max}} S F(S) dS = A \displaystyle \int_{S_{\rm thr}}^{S_{\rm max}} 
S dS \int_{S_{0, {\rm l}}}^{S_{0,{\rm h}}}
\displaystyle \frac{P_0(S_0)}{S_0} dS_0 P(a)\\
\dot{E}_{\rm true} = ~~~~ B f_b 
\displaystyle \int_{S_{0, \rm min}}^{S_{0, \rm max}} S_0 P_0(S_0) dS_0 
\, .
\end{cases}
\ee

In this case, 
the ratio ${\cal C}$ has a more complicated expression 
\be
\label{eq:Cratio-P0}
{\cal C} = \int_{a_{\rm thr}}^{a_{\rm max}} P(a) da \displaystyle \frac{\displaystyle \int P_0(S_0) dS_0}{\displaystyle \int S_0 P_0(S_0) dS_0} \displaystyle \frac{\displaystyle \int S P(S) dS}{\displaystyle \int P(S) dS} = \displaystyle  \frac{\langle S \rangle}{\langle S_0 \rangle}~\int_{a_{\rm thr}}^{a_{\rm max}} P(a) da 
\ee
which depends on the functional form of $P_0(S_0)$, as well as on $P(a)$ and the range of observable amplifications.~Note that, in deriving Eq.~\ref{eq:Cratio-P0}, we also used Eq.~\ref{eq:rateb-vs-rate} to express the ratio $A/B$.  

\section{General constraints on the local (unlensed) population}
\label{app:C}

The number of {\it unlensed} repeaters detectable at the CHIME sensitivity (Eq.~\ref{eq:ethrz}) is strongly dependent on the intrinsic energy per event ($E_0$) and the intrinsic burst rate per source (${\cal R}$), both of which may be decaying with time.~By comparing the most active repeaters with the few local sources that are known to repeat (e.g. SGR 1935+215 and FRB 20200120E), and with the estimated value of~${\cal R}$~in~Sec.~\ref{sec:results}, the presence of such evolutionary effects appear to be likely.~

Here we sketch this evolution as a simple power-law decay of the burst rate with a characteristic time $t_0$, such that
\be
\label{eq:burstdecay}
{\cal R} (t) = r_0 \left(1+ \displaystyle \frac{t}{t_0}\right)^{-\alpha} \, ,
\ee
where $r_0$ is the initial burst rate, and $t_0$ determines a plateau, during which ${\cal R}$ stays nearly constant before decaying $\propto t^{-\alpha}$.~Given~a local SN rate $\sim 10^{-4}$yr$^{-1}$Mpc$^{-1}$ and a 10\% magnetar fraction, the number of sources detectable as {\it unlensed} repeaters in a 1-yr long CHIME survey ($\sim$ 2 days per source) can be calculated using Eq.~\ref{eq:ethrz}.~We estimate this number to be less than 10 provided that $E_0 \lesssim 10^{36}$~erg, which limits the horizon for unlensed events, and that ${\cal R}(t) < 1$ day$^{-1}$ within $\sim 300/E^{3/2}_{0, 36}$ yrs of the source birth, where we normalized $E_0$ in units of $10^{36}$ erg. 

The more evolved members (age $> 300/E^{3/2}_{0,36}$ yrs) of the same population of nearby sources will produce {\it unlensed} one-off events in one year worth of CHIME data, which should provide at most a small contribution to the observed all-sky rate $\sim 500$~day$^{-1}$.

The emerging scenario, which was very simply outlined here, provides a consistent - though necessarily approximate - picture of the FRB source population.~With a characteristic $E_0 \lesssim 10^{36}$ erg, and a typical rate ${\cal R} < $ a few per year per source at an age $> 10^3$~yrs, the population of nearby sources here identified is perfectly in line with the observed properties of the two local FRBs, SGR 1935+215 
and FRB 20200120E.~Moreover, the youngest (age $\lesssim$ 100 yrs) and most active (${\cal R} \gtrsim 0.25$ hr$^{-1}$) members of that population are rare and only become apparent from larger distances, where some level of amplification is required for detection.

Finally, a more realistic expectation is that FRBs have a (broad) distribution of intrinsic energies, as opposed to a single $E_0$~value.\\ In particular, the most active repeaters may produce intrinsically more energetic bursts than the rest of the population, e.g.~due to a stronger interior B-field.~We 
note, though, that their
absence in the local universe would impose such repeaters to be outliers even in terms of their intrinsic properties:~in order to match the CHIME observations, we estimate that no more than 1\% of the young magnetar population may be characterized by E$_0 \gtrsim 2\times 10^{37}$ erg.~A more detailed population study is well beyond the scope of this work and is in the focus of a future publication.
\end{document}